\documentclass[12pt, a4paper]{article}

\usepackage[T1]{fontenc}
\usepackage{bigdelim}
\usepackage[utf8]{inputenc}
\usepackage{microtype}
\usepackage{lmodern}
\usepackage[lining]{ebgaramond} 
\usepackage[scaled=0.85]{beramono}
\usepackage{amssymb}
\usepackage[
top=2.8cm, bottom=2.8cm,
left=2.5cm,  right=2.5cm
]{geometry}
\usepackage{dsfont}
\usepackage{xcolor}
\usepackage{titlesec}
\usepackage{titling}
\usepackage{abstract}
\usepackage{booktabs}
\usepackage{amsmath, amssymb, amsthm}
\usepackage{graphicx}
\usepackage{hyperref}
\usepackage{fancyhdr}
\usepackage{tcolorbox}
\usepackage{setspace}
\usepackage{parskip}
\usepackage[round]{natbib}
\usepackage{enumitem}
\usepackage{float}
\usepackage{booktabs,array,longtable,multirow}
\usepackage{subcaption} 
\usepackage{caption}    

\definecolor{deepblue}{RGB}{18, 52, 104}
\definecolor{ruleblue}{RGB}{60, 100, 170}
\definecolor{lightblue}{RGB}{220, 230, 245}

\hypersetup{
	colorlinks  = true,
	linkcolor   = deepblue,
	citecolor   = deepblue,
	urlcolor    = ruleblue,
	pdftitle    = {Machine learning methods for survey sampling},
	pdfauthor   = {Dagdoug and Haziza (2026)},
}

\usepackage{algorithm}
\usepackage{algpseudocode}

\renewcommand{\headrulewidth}{0.4pt}
\renewcommand{\headrule}{\hbox to\headwidth{\color{ruleblue}\leaders\hrule height \headrulewidth\hfill}}
\fancypagestyle{plain}{%
	\fancyhf{}
	\fancyhead[R]{\small\color{deepblue}\thepage}
	\renewcommand{\headrulewidth}{0pt}
}

\titleformat{\section}
{\large\bfseries\color{deepblue}}
{\large\thesection.}{0.6em}{}

\titleformat{\subsection}
{\normalsize\bfseries\color{deepblue!80!black}}
{\thesubsection.}{0.5em}{}

\titleformat{\subsubsection}
{\normalsize\itshape\color{deepblue!70!black}}
{\normalsize\thesubsubsection.}{0.5em}{}

\titlespacing{\section}{0pt}{3.0ex plus .5ex}{1.8ex plus .3ex}
\titlespacing{\subsection}{0pt}{3.0ex plus .5ex}{1.5ex plus .3ex}
\titlespacing{\subsubsection}{0pt}{2.6ex plus .4ex}{1.2ex plus .2ex}

\tcbuselibrary{breakable, skins}
\newtcolorbox{abstractbox}{
	enhanced,
	breakable,
	colback     = lightblue,
	colframe    = deepblue,
	arc         = 3pt,
	boxrule     = 0.6pt,
	left        = 12pt,
	right       = 12pt,
	top         = 8pt,
	bottom      = 8pt,
	title       = {\textbf{\color{white}Abstract}},
	attach boxed title to top left = {yshift=-2mm, xshift=6pt},
	boxed title style = {
		colback   = deepblue,
		colframe  = deepblue,
		arc       = 2pt,
		boxrule   = 0pt,
	},
	fontupper   = \small\onehalfspacing,
}
\newtheorem{assumption}{Assumption}
\pretitle{%
	\begin{center}
		\vspace*{-0.5cm}
		{\color{deepblue}\rule{\linewidth}{1.2pt}}\\[0.5em]
		\LARGE\bfseries\color{deepblue}
	}
	\posttitle{%
		\\[0.1em]
		{\color{deepblue}\rule{\linewidth}{1.2pt}}
	\end{center}
	\vspace{0.2em}
}
\preauthor{\begin{center}\large\lineskip 0.75em\begin{tabular}[t]{c}}
		\postauthor{\end{tabular}\end{center}}
\predate{\begin{center}\small\itshape\color{gray}}
	\postdate{\end{center}\vspace{0.5em}}

\def\bx{\mathbf{x}}
\def\E{\mathds{E}}
\def\P{\mathds{P}}
\def\V{\mathds{V}}

\newcommand{\R}{\mathbb{R}}

\begin{document}
	
	\title{Machine learning methods for finite population parameter estimation in survey sampling}
	
\author{%
	\textbf{Mehdi Dagdoug}\thanks{Email: \texttt{mehdi.dagdoug@mcgill.ca}}\\[0.3em]
	{\small Department of Mathematics and Statistics}\\
	{\small McGill University, Montréal, Canada}
	\and
	\textbf{David Haziza}\thanks{Email: \texttt{dhaziza@uottawa.ca}}\\[0.3em]
	{\small Department of Mathematics and Statistics}\\
	{\small University of Ottawa, Ottawa, Canada}
}
	
	\date{}
	
	\maketitle
	\thispagestyle{plain}

	\begin{abstractbox}
	This pedagogical review examines the use of machine learning methods in finite-population inference for survey sampling, with an emphasis on design-based validity and statistical inference. While flexible prediction tools offer substantial gains in estimation accuracy, they also introduce important challenges, primarily due to the dependence between the fitted predictors and the sample. We focus on settings in which such predictions enter survey estimation through model-assisted estimation, item nonresponse imputation, and unit nonresponse adjustment. For model-assisted estimation and item nonresponse, we show how cross-fitting and Neyman-orthogonal estimating equations can adapt ideas from double/debiased machine learning to survey data, allowing the use of high-dimensional or nonparametric learners while preserving root-n consistency and asymptotic normality under suitable conditions. In contrast, for unit nonresponse, standard inverse-probability weighting remains outcome-agnostic and operationally attractive, but this same feature makes doubly robust and orthogonal constructions harder to deploy in official statistics. We also briefly discuss related developments in small area estimation and probability/non-probability data integration. Overall, the paper highlights both the promise of machine learning and the fundamental inferential challenges it raises for survey practice.
		

	\end{abstractbox}
	
	\vspace{1.2em}
	
	\section{INTRODUCTION}\label{Sec:INTRO}
In the last decade, machine learning (ML) methods have begun to attract considerable attention within national statistical offices as potential tools for enhancing the production of official statistics. Their emergence coincides with increasing availability of large and complex auxiliary data sources, advances in computational power, and sustained interest in modernising statistical systems. Although probability sampling and traditional inferential frameworks remain central to the mandate of national statistical offices, ML offers a flexible predictive layer that is capable of exploiting high-dimensional information, capturing nonlinear patterns, and generating accurate and robust predictions. In this paper, we use the term machine learning to refer broadly to a class of data-adaptive prediction methods that estimate an unknown regression or propensity function over a potentially high-dimensional class of functions, rather than by specifying a fixed finite-dimensional parametric model. Typical examples include tree-based methods, kernel and spline smoothers, and neural networks, among many others. Throughout, we use the term architecture to refer to a specific learning procedure together with its associated tuning parameters (or hyperparameters) (e.g., tree depth, regularization parameters, number of layers).\\

\noindent ML-based prediction enters survey estimation in several ways. These applications differ in their goals and assumptions, but they share a common structure: ML methods are used to produce fitted values, response propensities, or other predicted quantities that serve as building blocks for estimators of finite-population parameters, such as totals and means. First, model-assisted estimation (e.g., \cite{sarndal1992model, breidt2017model}) uses predictions based on auxiliary variables to improve efficiency while preserving design-based validity. A classical example is the generalized regression estimator, which combines model-based predictions with design weights. Second, missing data in surveys arise through either item nonresponse or unit nonresponse, and the two problems are usually treated differently. For item nonresponse, a common strategy is single imputation, where missing values are replaced by plausible values constructed from auxiliary information (e.g., \cite{chen2019recent}). For unit nonresponse, treatment typically relies on inverse probability weighting, where response propensities are modeled using auxiliary variables and incorporated into the survey weights, often together with calibration to known population totals (e.g., \cite{kalton2003weighting, kim2007nonresponse, haziza2017construction}). Third, small area estimation (e.g., \cite{rao2015small}) uses models and auxiliary information to produce estimates for domains where direct survey estimates are too unstable because of small sample sizes. Finally, the integration of probability and non-probability samples has received considerable attention in recent years (e.g., \cite{elliott2017inference, kim2019sampling, beaumont2020probability, chen2020doubly, rao2021making}). This interest is partly driven by the growing availability of alternative data sources, such as administrative records and web panels, and by the increasing difficulties faced by traditional probability surveys, including declining response rates and rising data-collection costs. The goal is to combine the strengths of both sources: probability samples provide a principled basis for inference, while non-probability samples may offer larger sample sizes, better timeliness, or more detailed information. \\

\noindent \noindent In survey sampling, machine learning methods are increasingly used as flexible prediction tools in the construction of estimators. Unlike classical parametric models, ML methods are highly adaptive and are often trained on the same data used for estimation, creating dependence between the fitted predictor and the sample. This dependence can affect bias, variance estimation, and the construction of valid confidence intervals. Three related difficulties arise. First, many ML predictors converge more slowly than $1/\sqrt{n}$, with rates depending on the smoothness of the underlying regression function and on the dimension of the covariate space. When such predictions are incorporated into survey estimators—through imputation, model-assisted estimation, or propensity modelling—the resulting estimators may fail to satisfy the usual $\sqrt{n}$ asymptotics. Second, these estimators are often not Neyman-orthogonal, so estimation error in the fitted nuisance functions can contribute at the first order. In classical survey settings, this additional variability can often be handled using linearization or analytic approximations, but this becomes more difficult when nuisance functions are estimated using flexible and non-smooth ML procedures. In such cases, resampling methods such as the bootstrap may also fail to reproduce the sampling distribution accurately. Third, when prediction error dominates sampling variability, asymptotic normality may fail, making standard large-sample inference unreliable. In this paper, we argue that these issues can be traced back to two main sources: the way predictions enter some classical estimators, and the dependence between predictions and the sample. This perspective helps explain the behavior of ML methods in several survey sampling settings.\\

\noindent A growing body of work, outside of the survey sampling literature, (e.g., \cite{chernozhukov2018double, kennedy2024semiparametric, ahrens2025introduction}) addresses these challenges through the use of doubly robust or Neyman-orthogonal estimators. Traditionally, doubly robust estimators were introduced to protect against model misspecification: they remain consistent if either the outcome model or the propensity model is correctly specified, though not necessarily both (e.g., \cite{robins1994estimation, kang2007demystifying, cao2009improving, kim2014doubly}). A less widely emphasized property is that, when appropriately constructed, doubly robust estimators can also achieve $\sqrt{n}$-consistency even when nuisance functions, such as response propensities or outcome regressions, converge more slowly than $1/\sqrt{n}$. This property relies on orthogonal estimating equations, which remove the first-order sensitivity of the estimator to nuisance estimation errors, so that these errors affect the estimator only through higher-order terms. A second key ingredient is cross-fitting (e.g., \cite{chernozhukov2018double, newey2018cross, lu2025conditional}), in which prediction models are trained on one subset of the data and evaluated on another. This reduces overfitting and allows the orthogonal estimating equations to behave as if the nuisance functions had been estimated on an independent sample, helping to restore asymptotic normality and standard Wald-type inference.\\

\noindent Our goal is pedagogical rather than methodological: we review how double/debiased machine learning (DML) ideas can be adapted to survey data collected under possibly complex sampling designs. After introducing the basic setup in Section~2, Section~3 reviews model-assisted estimation and shows how orthogonalization can help recover design-based inference when prediction models converge more slowly than $n^{-1/2}$. Section~4 considers imputation for item nonresponse and discusses how debiasing principles apply when imputed values are constructed using flexible ML models. Section~5 turns to unit nonresponse, where inverse probability weighting is still the standard approach in official statistics. Because nonresponse-adjusted weights are typically used for many survey variables rather than a single target parameter, DML is harder to use in this setting; we therefore discuss the main difficulties and open challenges. Finally, Section~6 briefly reviews two related areas in which ML is increasingly used: small area estimation and the integration of probability and non-probability samples.

	\section{THE SETUP: FINITE POPULATION ESTIMATION}

Consider a finite population $U=\{1,\ldots,N\}$ of $N$ units. For example, $U$ may represent the population of Canadian households or the population of active businesses in Canada. We are interested in estimating the finite-population mean of a survey variable $Y$,
\[
\mu \;=\; \dfrac{1}{N}\sum_{k\in U} y_k,
\]
where $y_k$ denotes the $y$-value associated with unit $k\in U$. A sample $S$ of (possibly random) size $n$ is selected according to a sampling design $\mathcal{F}(\mathbf{I}\mid \mathbf{Z})$, where $\mathbf{Z}$ denotes the matrix of design information available prior to sampling for all population units, and $\mathbf{I}=(I_1,\ldots,I_k,\ldots,I_N)^\top$ is the $N$–vector of sample selection indicators, with $I_k=1$ if $k\in S$ and $I_k=0$ otherwise. The first– and second–order inclusion probabilities are defined as
\[
\pi_k \;=\; \P(k \in S\mid \mathbf{Z})= \E(I_k\mid \mathbf{Z}), 
\qquad 
\pi_{k\ell} \;=\; \P(k, \ell \in S\mid \mathbf{Z})= \E(I_k I_\ell \mid \mathbf{Z}), \qquad (k\neq \ell \in U).
\]
When summations include diagonal terms, we use the convention $\pi_{kk}=\pi_k$.	\\

\noindent   Commonly encountered sampling designs include:
\emph{(i) Simple random sampling without replacement (SRSWOR)}, in which a fixed-size sample of size $n$ is drawn uniformly from the ${N\choose n}$ subsets of $U$, yielding equal first-order inclusion probabilities $\pi_k=n/N$ for all $k\in U$ and second-order inclusion probabilities $\pi_{k\ell}=n(n-1)/\{N(N-1)\}$ for $k\neq \ell$;
\emph{(ii) Inclusion–probability–proportional–to–size (IPPS) sampling without replacement}, where a strictly positive size measure $z_k$  determines the inclusion probabilities:
\[
\pi_k \;=\; \frac{n\,z_k}{\sum_{j\in U} z_j}, \qquad 0 < \pi_k \le 1, \quad k\in U,
\]
with a fixed sample size $n$. The second-order inclusion probabilities $\pi_{k\ell}$ depend on the specific IPPS algorithm used to implement the design (e.g., \cite{tille2006sampling});
\emph{(iii) Poisson sampling}, in which each unit $k$ in the population is selected independently through a Bernoulli trial with first-order inclusion probability $\pi_k$, so that $\pi_{k\ell} = \E(I_k I_\ell \mid \mathbf{Z})=\E(I_k \mid \mathbf{Z})\E(I_\ell \mid \mathbf{Z})=\pi_k \pi_\ell$ for $k \neq \ell$. Under this design, the sample size is random and follows a Poisson–binomial distribution, with expected sample size $\E(|S|\mid \mathbf{Z}) = \sum_{k \in U} \pi_k$; and \emph{(iv) Stratified sampling}, where $U$ is partitioned into strata $U_1,\ldots,U_H$ (e.g., by region or industry) with sizes $N_h$. Typically, SRSWOR of size $n_h$ is conducted independently within each stratum, giving $\pi_k=n_h/N_h$ for $k\in U_h$ and $\pi_{k\ell}=n_h(n_h-1)/\{N_h(N_h-1)\}$ when $k\neq \ell$ belong to the same stratum $U_h$, and $\pi_{k\ell}=(n_h/N_h)(n_{h'}/N_{h'})$ across strata $U_h$ and $U_{h^{'}}$, with $h \neq h'$.\\

\noindent We now introduce a set of conditions on the sampling design that will be required for subsequent results.
\begin{assumption}[First-order design positivity]\label{ass:design_positivity}
	All population units have a nonzero probability of selection under the sampling design, that is,
	$\pi_k > 0$ for all $ k \in U$.
\end{assumption}

\begin{assumption}[Second-order design positivity]\label{ass:design_second_order_positivity}
	All distinct pairs of population units have a positive second-order inclusion probability, that is,
	$\pi_{k\ell} > 0$ for all  $k \neq \ell,\; k,\ell \in U$.
\end{assumption}
\noindent In practice, Assumption 1 fails in settings where some units have zero chance of being included in the sample. This occurs, for instance, under undercoverage, where members of the population are missing from the sampling frame (e.g., households without internet in a web survey), or in cut-off sampling, where units below or above a threshold are deliberately excluded. In both cases, some units satisfy $\pi_k=0$, violating first-order design positivity. Assumption 2 is satisfied under most standard sampling designs. A notable exception is systematic sampling, where many pairs of units can never be jointly selected, leading to $\pi_{k\ell} = 0$ for some $k \neq \ell$.\\

\noindent A common estimator of the population mean is the Horvitz–Thompson estimator 
\begin{equation}\label{eq:HT_mean}
	\widehat{\mu}_{\mathrm{HT}} \;=\; \frac{1}{N}\sum_{k \in S} w_k\,y_k,
	\qquad w_k:=\pi_k^{-1}.
\end{equation}
The quantity $w_k$ is commonly referred to as the sampling weight of unit $k$. Intuitively, it represents the number of population units that the sampled unit $k$ stands for under the sampling design. \\

\noindent To study the bias and variance of the Horvitz–Thompson estimator under ideal conditions, we work in the design-based paradigm, which is standard in survey sampling. In this framework, the finite-population values $\mathbf{y}=(y_1,\ldots,y_N)^{\top}$ and the design information $\mathbf{Z}$ are treated as fixed quantities, while randomness arises solely from the sampling design $\mathcal{F}(\mathbf{I} \mid \mathbf{Z})$. Therefore, when taking expectations or variances, all quantities except the sample selection indicators $\mathbf{I}$ are regarded as fixed. To reflect this, we denote design expectations and variances by $\E_d(\cdot)$ and $\V_d(\cdot)$, respectively.\\

\noindent Under Assumption~1, $\widehat{\mu}_{\mathrm{HT}}$ is design-unbiased or $d$-unbiased for $\mu$, i.e., $\E_d\!\left(\widehat{\mu}_{\mathrm{HT}}\right) = \mu$. Its design variance is
\begin{equation}\label{eq:Var_HT_mean}
	\V_d\!\left(\widehat{\mu}_{\mathrm{HT}}\right)
	\;=\;
	\frac{1}{N^{2}}
	\sum_{k\in U}\sum_{\ell\in U}
	\Delta_{k\ell}\;
	\frac{y_k}{\pi_k}\;\frac{y_\ell}{\pi_\ell},
\end{equation}
where $\Delta_{k\ell} := \mbox{Cov}(I_k, I_\ell\mid \mathbf{Z}) = \pi_{k\ell}-\pi_k\pi_\ell$. When both Assumptions~1 and~2 hold, an unbiased estimator of \eqref{eq:Var_HT_mean} is
\begin{equation}\label{eq:Vhat_HT_mean}
	\widehat{\V}\!\left(\widehat{\mu}_{\mathrm{HT}}\right)
	\;=\;
	\frac{1}{N^{2}}
	\sum_{k\in S}\sum_{\ell\in S}
	\frac{\Delta_{k\ell}}{\pi_{k\ell}}\;
	\frac{y_k}{\pi_k}\;\frac{y_\ell}{\pi_\ell}.
\end{equation}
That is, $
\E_d\!\{\widehat{\V}\!(\widehat{\mu}_{\mathrm{HT}})\}
=
\V_d\!(\widehat{\mu}_{\mathrm{HT}}).
$ 
Under appropriate higher-order assumptions on the sampling design and the survey variable, it can also be shown that $\widehat{\V}\!(\widehat{\mu}_{\mathrm{HT}})$ is consistent for $	\V_d\! (\widehat{\mu}_{\mathrm{HT}})$. 	 Beyond unbiased point estimation and variance estimation, design-based inference often relies on large-sample normal approximations. Under suitable regularity conditions on the sampling design, central limit theorems have been established for Horvitz--Thompson estimators and related classes of estimators. These results justify the construction of Wald-type confidence intervals based on $\widehat{\V}(\widehat{\mu}_{\mathrm{HT}})$. Classical references include \cite{hajek1960limiting, hajek1964asymptotic, krewski1981inference, rosen1997asymptotic, chen2007asymptotic}, among others.  An approximate $(1-\alpha)$ Wald-type confidence interval for the population mean $\mu$ is given by
\[
\widehat{\mu}_{\mathrm{HT}} \;\pm\; z_{1-\alpha/2}\,
\big\{\widehat{\V}(\widehat{\mu}_{\mathrm{HT}})\big\}^{1/2},
\]
where $z_{1-\alpha/2}$ denotes the $(1-\alpha/2)$ quantile of the standard normal distribution and $\widehat{\V}(\widehat{\mu}_{\mathrm{HT}})$ is given by \eqref{eq:Vhat_HT_mean}

\section{MODEL-ASSISTED ESTIMATION}

While the Horvitz--Thompson estimator $\widehat{\mu}_{\mathrm{HT}}$ is $d$-unbiased
under Assumption~1, it can be inefficient when
the survey variable $Y$ is strongly related to auxiliary variables
$\mathbf{x}$, because the estimator does not exploit this information.
Model-assisted estimation addresses this limitation by explicitly
incorporating auxiliary variables through a predictive statistical model,
with the objective of improving efficiency relative to the
Horvitz--Thompson estimator. The key feature of
the model-assisted approach is that the finite-population values
$\mathbf{y}=(y_1,\ldots,y_N)^\top$ are treated as fixed quantities, and
inference is evaluated with respect to the sampling design. We emphasize that the working
model is not assumed to be correctly specified; rather, it is used as a
device to construct estimators that exploit the association between $Y$
and auxiliary variables while retaining a design-based validity and interpretation.\\

\noindent The roots of model-assisted estimation trace back to early work on
regression estimation in survey sampling. Classical references include \cite{cochran1942sampling, cochran1977sampling} on regression estimators, and
\cite{cassel1976some} on regression and generalized
regression ideas. Important  contributions to the
model-assisted paradigm were provided by \cite{sarndal1980pi} and by
\cite{robinson1983asymptotic}, who clarified the role of working models
in design-based inference. These developments were synthesized in the
monograph of \cite{sarndal1992model}, which formalized the
model-assisted approach. A comprehensive modern review of model-assisted estimation and its
extensions can be found in \cite{breidt2017model}.

\subsection{ORACLE MODEL-ASSISTED ESTIMATION}\label{Sec:OR:MA}

Let $\mathbf{X}=(\mathbf{x}_1^\top,\ldots,\mathbf{x}_N^\top)^\top$ denote the
$N\times p$ matrix of auxiliary variables, assumed to be available at the
estimation stage. Let $\mathbf{x}_k$ denote the vector of auxiliary
variables associated with unit $k\in U$. In model-assisted estimation, we consider a working prediction architecture (e.g., linear regression, splines, random forests, neural networks) aimed at estimating  the regression function $m(\mathbf{x}_k)=\E(y_k\mid \mathbf{x}_k)$. \\

\noindent We denote by $m_N(\cdot)$ a target prediction rule associated with the
chosen learning strategy. This rule should be viewed as the stable prediction
target toward which the fitted rule $\widehat m(\cdot)$, trained from the
sample, is expected to converge. The subscript $N$ emphasizes that this target
may depend on the finite population size and on the learning strategy. Importantly, $m_N(\cdot)$ is treated as fixed under the sampling design. It is not assumed to be the true regression function $m(\mathbf x)=\E(y\mid \mathbf x)$, nor is it necessarily the prediction rule
that would be obtained by training the same algorithm on the full finite
population. In classical model-assisted estimation, $m_N(\cdot)$ is often naturally
interpreted as a population-level fitted rule. For example, under a linear
working model, one usually takes
\[
m_N(\mathbf x_k)=\mathbf x_k^\top \mathbf B_N,
\qquad
\mathbf B_N
=
\left(
\sum_{\ell\in U}\mathbf x_\ell\mathbf x_\ell^\top
\right)^{-1}
\sum_{\ell\in U}\mathbf x_\ell y_\ell.
\]
In that case, the target prediction rule coincides with the fitted rule that
would be obtained using the full finite population. For more flexible machine
learning procedures, however, the architecture, tuning parameters, or training
procedure may depend on the sample size or on the way the learner is trained.
In such cases, the full-population fit is not always the most appropriate
theoretical benchmark. We therefore use $m_N(\cdot)$ more generally to denote
a fixed oracle target for the fitted prediction rule.\\

\noindent If $m_N(\mathbf x_k)$ were known for all $k\in U$, an oracle
model-assisted estimator, often referred to as a difference estimator, of the
finite population mean $\mu$ is
\begin{equation}\label{eq:oracle_MA_mN}
	\widehat{\mu}_{\mathrm{MA}}(m_N)
	=
	\frac{1}{N}
	\left[
	\sum_{k\in U} m_N(\mathbf{x}_k)
	+
	\sum_{k\in S} w_k\,
	\{y_k-m_N(\mathbf{x}_k)\}
	\right].
\end{equation}
The notation $\widehat{\mu}_{\mathrm{MA}}(m_N)$ emphasizes that the estimator
is a functional of the oracle target rule $m_N(\cdot)$. The first term corresponds to the finite-population mean of the predicted
values and, taken alone, is generally design-biased for $\mu$. The second
term is a Horvitz--Thompson estimator of the finite-population mean of the
oracle residuals
$
e_{Nk}=y_k-m_N(\mathbf{x}_k),
$
and acts as a design-bias correction for the prediction error induced by
$m_N(\cdot)$. As a result, $\widehat{\mu}_{\mathrm{MA}}(m_N)$ is
$d$-unbiased regardless of the predictive quality of $m_N(\cdot)$. Indeed,
under Assumption~\eqref{ass:design_positivity},
\[
\E_d\!\left\{\widehat{\mu}_{\mathrm{MA}}(m_N)\right\}
=
\frac{1}{N}\sum_{k\in U} m_N(\mathbf{x}_k)
+
\frac{1}{N}\sum_{k\in U} \E_d(I_k w_k)\,
\{y_k-m_N(\mathbf{x}_k)\}
=
\mu,
\]
since $\E_d(I_k w_k)=1$ for all $k\in U$. Thus, design-based validity does
not rely on $m_N(\cdot)$ being a correct model for $m(\cdot)$. When the
target rule $m_N(\cdot)$ captures a substantial portion of the systematic
variation in $Y$, the residuals $y_k-m_N(\mathbf{x}_k)$ tend to be smaller
and the resulting estimator may be more efficient than
$\widehat{\mu}_{\mathrm{HT}}$. Under standard regularity conditions,
$\widehat{\mu}_{\mathrm{MA}}(m_N)$ is root-$n$ consistent.\\
\noindent
We now turn to the design variance of $\widehat{\mu}_{\mathrm{MA}}(m_N)$. Since the first
term in equation \eqref{eq:oracle_MA_mN} is nonrandom under the sampling design, the
design variance depends only on the second term, leading to
\begin{equation}\label{eq:V_ma_or_mN}
	\V_d\!\left\{\widehat{\mu}_{\mathrm{MA}}(m_N)\right\}
	=
	\frac{1}{N^2}
	\sum_{k\in U}\sum_{\ell\in U}
	\Delta_{k\ell}\;
	\frac{y_k-m_N(\mathbf{x}_k)}{\pi_k}\;
	\frac{y_\ell-m_N(\mathbf{x}_\ell)}{\pi_\ell}.
\end{equation}
Expression~\eqref{eq:V_ma_or_mN} shows that the design variance is driven by
the  census residuals $y_k-m_N(\mathbf{x}_k)$ and is therefore small when the chosen
architecture yields accurate population-level fits. Under Assumptions~\eqref{ass:design_positivity} and
\eqref{ass:design_second_order_positivity}, a $d$-unbiased  and consistent estimator of
$	\V_d\!\left\{\widehat{\mu}_{\mathrm{MA}}(m_N)\right\}$ is
\begin{equation}\label{eq:Vhat_oracle_MA_mN}
	\widehat{\V}\!\left\{\widehat{\mu}_{\mathrm{MA}}(m_N)\right\}
	=
	\frac{1}{N^2}
	\sum_{k\in S}\sum_{\ell\in S}
	\frac{\Delta_{k\ell}}{\pi_{k\ell}}\;
	\frac{y_k-m_N(\mathbf{x}_k)}{\pi_k}\;
	\frac{y_\ell-m_N(\mathbf{x}_\ell)}{\pi_\ell}.
\end{equation}
That is,
\[
\E_d\!\left[
\widehat{\V}\!\left\{\widehat{\mu}_{\mathrm{MA}}(m_N)\right\}
\right]
=
\V_d\!\left\{\widehat{\mu}_{\mathrm{MA}}(m_N)\right\}.
\]
Moreover, since the estimator in \eqref{eq:oracle_MA_mN} is a Horvitz--Thompson--type estimator
applied to the fixed residuals $y_k-m_N(\mathbf{x}_k)$, it inherits a central
limit theorem under similar regularity conditions used for Horvitz--Thompson
estimators. As a result,
\[
\left[\V_d\!\left\{\widehat{\mu}_{\mathrm{MA}}(m_N)\right\}\right]^{-1/2}
\big\{\widehat{\mu}_{\mathrm{MA}}(m_N)-\mu\big\}
\xrightarrow[n \to \infty]{\mathcal{L}}
\mathcal{N}\!\left(0,1\right),
\]
which, together with \eqref{eq:Vhat_oracle_MA_mN}, yields asymptotically valid
Wald-type confidence intervals for $\mu$.

\subsection{FEASIBLE MODEL-ASSISTED ESTIMATION}
In practice, the oracle target rule $m_N$ is unknown and must be approximated from the sample.
Let $\widehat m(\cdot)$ denote an estimator of $m_N(\cdot)$
constructed using the observed sample $S$. Replacing $m_N$ by $\widehat m$ in expression
\eqref{eq:oracle_MA_mN} yields the feasible model-assisted estimator
\begin{equation}\label{eq:MA_feasible}
	\widehat{\mu}_{\mathrm{MA}}(\widehat m)
	=
	\frac{1}{N}
	\left[
	\sum_{k\in U} \widehat m(\mathbf{x}_k)
	+
	\sum_{k\in S} w_k\,
	\{y_k-\widehat m(\mathbf{x}_k)\}
	\right].
\end{equation}
Estimator~\eqref{eq:MA_feasible} remains within the model-assisted framework:
predictions are used to improve efficiency, while inference is evaluated
with respect to the sampling design.\\

Insight into its behavior can be gained by
decomposing it into the oracle estimator based on $m_N$ and a remainder term
that captures the effect of estimating the population-level fitted rule:
\begin{eqnarray}\label{eq:MA_decomposition}
	\widehat{\mu}_{\mathrm{MA}}(\widehat m)
	&=&
	\widehat{\mu}_{\mathrm{MA}}(m_N)
	+
	\frac{1}{N}
	\sum_{k\in U}
	\left(1-\frac{I_k}{\pi_k}\right)
	\{\widehat m(\mathbf{x}_k)-m_N(\mathbf{x}_k)\}\nonumber \\
	& :=& \widehat{\mu}_{\mathrm{MA}}(m_N)  + 	R_N.
\end{eqnarray}
As discussed in Section~\eqref{Sec:OR:MA}, the oracle component
$\widehat{\mu}_{\mathrm{MA}}(m_N)$ fluctuates around $\mu$ at the $1/\sqrt{n}$ scale. Therefore, for the feasible estimator to
inherit the same first-order behavior, the remainder term in
\eqref{eq:MA_decomposition} must be asymptotically negligible, i.e., $R_N = o_\P (n^{-1/2})$.\\

\noindent Model-assisted estimators may be viewed as solutions of finite population estimating equations that satisfy a Neyman orthogonality property; see Section S.1 of the Supplementary Material. To show that $R_N = o_\P(n^{-1/2})$, it is sufficient, for instance, to establish that $N \E_d[R_N^2] = o(1)$. Using the notation $\alpha_k := 1 - I_k/\pi_k$ for $k \in U$, we can write
\begin{align} \label{eq:maProof}
	N \E_d[R_N^2] 
	&= \frac{1}{N} \sum_{k \in U} \E_d \left[\alpha_k^2 (\widehat m(\bx_k) - m_N(\bx_k))^2 \right]\nonumber \\
	&\quad + \frac{1}{N} \sum_{k \in U} \sum_{\substack{\ell \in U \\ \ell \neq k}} \E_d \left[\alpha_k \alpha_\ell \{\widehat m(\bx_k) - m_N(\bx_k)\} \{\widehat m(\bx_\ell) - m_N(\bx_\ell)\} \right].
\end{align}
The first term on the right-hand side of \eqref{eq:maProof} converges to $0$ provided that
\begin{equation}\label{Cond_MA}
	N^{-1}\sum_{k\in U} \E_d \left[(\widehat m(\bx_k) - m_N(\bx_k))^2 \right] \to 0.
\end{equation}
The second term is more delicate due to the dependence between $\alpha_k$ and $\alpha_\ell$ for $k \neq \ell$. For many common sampling designs, however, we have
\[
\max_{k \neq \ell \in U} \left| \E_d[\alpha_k \alpha_\ell] \right| = O(n^{-1}).
\]
The key step is to approximate
\begin{align} 
	\E_d &\left[\alpha_k \alpha_\ell (\widehat m(\bx_k) - m_N(\bx_k)) (\widehat m(\bx_\ell) - m_N(\bx_\ell)) \right]\nonumber\\
	&\approx 
	\E_d \left[	(\widehat m(\bx_k) - m_N(\bx_k)) (\widehat m(\bx_\ell) - m_N(\bx_\ell))\right] \, \E_d[\alpha_k \alpha_\ell],\label{eq:keyEq}
\end{align}
which would effectively decouple the learning component from the sampling design. This approximation would hold exactly if $\widehat m$ were independent of the sample $S$. In that case, the weak dependence structure of $\E_d[\alpha_k \alpha_\ell]$ could be exploited, and the second term would also be negligible. Therefore, provided that Condition \eqref{Cond_MA} holds,
it follows that $R_N = o_\P(n^{-1/2})$. Consequently, the feasible model-assisted estimator inherits the root-$n$ consistency and asymptotic normality of the corresponding oracle estimator.\\

\noindent When $\widehat{m}$ depends on the sample in a relatively smooth and structured way, as in linear or generalized linear models (e.g., \cite{lehtonen1998logistic,firth1998robust}), it is often possible to linearize $\widehat{m}$ as a function of the sampling indicators. This linearization can then be exploited to establish an approximate independence between the estimation of $\widehat{m}$ and the sampling design, thereby justifying approximations of the form \eqref{eq:keyEq}. In contrast, for more flexible nonparametric or machine learning methods, $\widehat{m}$ typically depends on the sample in a complex and highly nonlinear manner. In such cases, a linearization argument is generally not available, and approximations such as \eqref{eq:keyEq} need not hold. Nevertheless, model-assisted estimation based on flexible nonparametric and machine-learning methods has been widely studied in the survey sampling literature. These approaches include kernel and local polynomial regression \citep{breidt2000local}, additive models \citep{opsomer2007model, wang2011nonparametric}, penalized splines \cite{breidt2005model}, regression splines \cite{goga2005reduction}, neural networks \cite{montanari2005nonparametric}, $K$-nearest neighbour methods \citep{baffetta2009design}, regression trees \citep{mcconville2019automated}, and random forests \citep{dagdoug2023model}. \cite{sanguiaosande2021design} proposed a subsampling Rao--Blackwellization procedure that allows flexible machine-learning algorithms to be used for model-assisted estimation while preserving exact design unbiasedness.\\

\noindent The same dependence between $\widehat m$ and the sample $S$ also affects variance estimation. A natural design-based variance estimator replaces the census residuals $y_k-m_N(\mathbf{x}_k)$ by the sample residuals $e_k=y_k-\widehat m(\mathbf{x}_k)$. When $\widehat m(\cdot)$ is highly adaptive, it may overfit the realized sample, artificially reducing the within-sample variability of $e_k$ and yielding downward-biased variance estimates. Cross-fitting provides a remedy: by separating the data used to train $\widehat m(\cdot)$ from the data used to evaluate residuals and construct the estimator, it weakens---and in some cases, such as Poisson sampling (see Section~\eqref{Sec:POI_MA}), eliminates---this dependence.\\

\subsection{Cross-fitted model-assisted estimation under Poisson sampling}
\label{Sec:POI_MA}

In this section, we discuss feasible model-assisted estimation under Poisson
sampling. The Poisson case is treated here in some detail because of its
analytical simplicity and its ability to clarify the role of cross-fitting.
In particular, the independence of the sample selection indicators $I_k$
greatly simplifies the arguments for design-unbiasedness and first-order
asymptotic properties. Other sampling designs are discussed in
Section~\eqref{Cross_Other_sam}\\

\begin{center}
	\begin{figure}[h!]
		\centering
		\resizebox{1\linewidth}{!}{
			\includegraphics[width=0.8\textwidth]{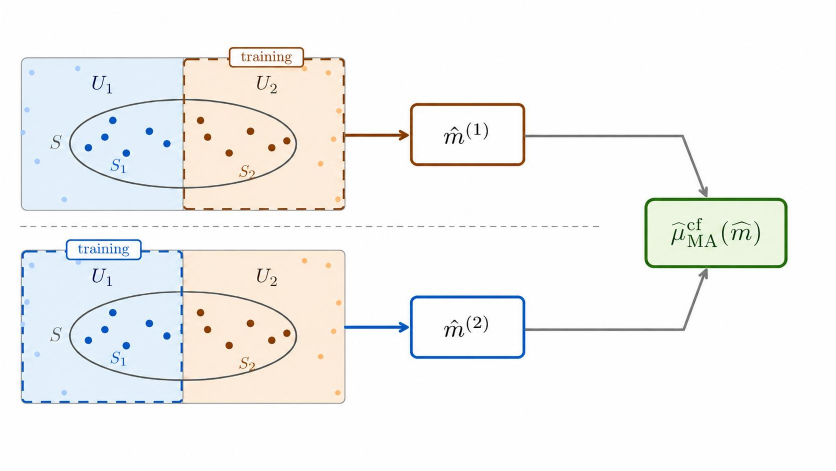}
		}
		\caption{Cross-fitting procedure for $K=2$.}	\label{Fig1}
	\end{figure}	
\end{center}

\noindent
Fix an integer $K\ge 2$ and partition the population $U$ into $K$ disjoint
folds $U_1,\ldots,U_K$; see Figure \ref{Fig1}. Let $v(k)\in\{1,\ldots,K\}$ denote the index of the
fold containing unit $k \in U$. The partition is constructed prior to sampling by
randomly dividing $U$ into $K$ disjoint subsets (for example, by sequential
simple random sampling without replacement from $U$) and is treated as fixed
under the sampling design. For each fold $v$, construct a prediction rule
$\widehat m^{(-v)}(\cdot)$ using only the sample observations belonging to
units in the complement $U_{-v}=U\setminus U_v$. The cross-fitted prediction
for unit $k$ is then defined as
$\widehat m^{(-v(k))}(\mathbf{x}_k)$. By construction,
$\widehat m^{(-v(k))}(\mathbf{x}_k)$ is independent of $I_k$, under Poisson
sampling. The cross-fitted feasible model-assisted estimator of the population mean is
defined as
\begin{equation}\label{eq:MA_crossfit}
	\widehat{\mu}_{\mathrm{MA}}^{\mathrm{cf}}(\widehat m)
	=
	\frac{1}{N}\sum_{k\in U}
	\left\{
	\widehat m^{(-v(k))}(\mathbf{x}_k)
	+
	\frac{I_k}{\pi_k}
	\big(y_k-\widehat m^{(-v(k))}(\mathbf{x}_k)\big)
	\right\}.
\end{equation}
\noindent Estimator~\eqref{eq:MA_crossfit} has the same algebraic structure as the
oracle estimator $\widehat{\mu}_{\mathrm{MA}}(m_N)$ introduced in
Section~\eqref{Sec:OR:MA}, with the population-level fitted rule $m_N(\cdot)$
replaced by cross-fitted predictions. The independence of $I_k$ and
$\widehat m^{(-v(k))}(\mathbf{x}_k)$ under Poisson sampling makes
design-unbiasedness immediate. Indeed, conditioning on
$\widehat m^{(-v(k))}(\mathbf{x}_k)$,
\begin{equation}\label{eq:cfUnbiased}
	\E_d\!\left[
	\widehat m^{(-v(k))}(\mathbf{x}_k)
	+
	\frac{I_k}{\pi_k}
	\big(y_k-\widehat m^{(-v(k))}(\mathbf{x}_k)\big)
	\;\middle|\;
	\widehat m^{(-v(k))}(\mathbf{x}_k)
	\right]
	=
	y_k.
\end{equation}
Summing over $k\in U$ yields
\[
\E_d\!\left\{\widehat{\mu}_{\mathrm{MA}}^{\mathrm{cf}}(\widehat m)\right\}
=
\mu.
\]
Importantly, this exact design-unbiasedness holds regardless of the quality
or correctness of the regression estimator $\widehat m(\cdot)$. This argument hinges critically on the independence
between $\widehat m^{(-v(k))}(\mathbf{x}_k)$ and $I_k$. Without cross-fitting,
the prediction $\widehat m(\mathbf{x}_k)$ is trained using the same realized
sample that determines $I_k$, and therefore depends on $I_k$ in a nontrivial
way; in that case,
$\E_d(I_k/\pi_k\mid \widehat m(\mathbf{x}_k))\neq 1,$ in general.\\

\noindent
It is useful to express the cross-fitted model-assisted estimator as
\begin{equation}\label{eq:MA_cf_decomposition}
	\widehat{\mu}_{\mathrm{MA}}^{\mathrm{cf}}(\widehat m)
	=
	\widehat{\mu}_{\mathrm{MA}}(m_N)
	+ \sum_{j=1}^K \dfrac{1}{N} \sum_{k\in U_j} \alpha_k\left( \widehat{m}^{(-j)}(\bx_k) - m_N(\bx_k) \right) := 	\widehat{\mu}_{\mathrm{MA}}(m_N)
	+ \sum_{j=1}^K R_j.
\end{equation}
The first term on the right-hand side is the oracle model-assisted estimator
discussed in Section~\eqref{Sec:OR:MA}. The second term captures the effect of
estimating the population-level fitted rule $m_N(\cdot)$. Under Poisson sampling, the variables
$\{\alpha_k:k\in U\}$ are independent with mean zero. For each $j=1, \ldots, K$, conditional on the construction of
$\widehat m^{(-j)}(\cdot)$, which is a measurable function of $(I_i)_{i\in U_{-j}}$, thus independent of $I_k/\pi_k-1$ for $k \in U_j$, the design variance of the second term on the
right-hand side of \eqref{eq:MA_cf_decomposition} is bounded by
\[
\V_d \left( R_j \big \rvert \widehat m^{(-j)} \right)\leqslant \frac{1}{N}\sum_{k\in U}\frac{1-\pi_k}{\pi_k}
\big\{\widehat m^{(-v(k))}(\mathbf{x}_k)-m_N(\mathbf{x}_k)\big\}^2 \times \dfrac{C}{N},
\]
for some constant $C>0$. Since $R_j$ has conditional mean zero, under appropriate regularity conditions on the sampling design, if
\begin{equation}\label{MSE_pred}
	\lim_{n \to \infty }\E_d \left[ \dfrac{	1}{N}\sum_{k\in U}
	\big\{\widehat m^{(-v(k))}(\mathbf{x}_k)-m_N(\mathbf{x}_k)\big\}^2
	\right] = 0,
\end{equation}
then the second term in expression \eqref{eq:MA_cf_decomposition} is
$o_{\P}(n^{-1/2})$, and hence
\[
\sqrt{n}\Big(
\widehat{\mu}_{\mathrm{MA}}^{\mathrm{cf}}(\widehat m)
-
\widehat{\mu}_{\mathrm{MA}}(m_N)
\Big)
=
o_{\P}(1).
\]
Therefore,
$\widehat{\mu}_{\mathrm{MA}}^{\mathrm{cf}}(\widehat m)$ is asymptotically
equivalent to the oracle estimator
$\widehat{\mu}_{\mathrm{MA}}(m_N)$ and inherits its first-order asymptotic
properties, including root-$n$ consistency and a central limit theorem. It is worth pointing out that this asymptotic equivalence does not require any specific
rate of convergence for $\widehat m(\cdot)$.
It suffices that \eqref{MSE_pred} holds, i.e., the cross-fitted predictions are
mean-square consistent for the population-level fitted rule $m_N(\cdot)$.
In particular, $m_N(\cdot)$ need not coincide with the true regression function
$m(\cdot)$.	Moreover, since $$ \V_d \left(\widehat{\mu}_{MA}(m_N)\right)= \dfrac{1}{N^2}\sum_{k\in U}\dfrac{1-\pi_k}{\pi_k} \left(y_k-m_N(\bx_k)\right)^2$$ in Poisson sampling, a natural variance estimator of $\widehat{\mu}_{\mathrm{MA}}^{\mathrm{cf}}(\widehat m)$ is given by 
$$ \widehat{\V}_d \left(\widehat{\mu}_{\mathrm{MA}}^{\mathrm{cf}}(\widehat{m})\right)= \dfrac{1}{N^2}\sum_{k\in S} \dfrac{1-\pi_k}{\pi_k^2} \left(y_k-\widehat{m}^{(-v(k))}(\bx_k)\right)^2,$$ using the honest residuals $y_k-\widehat{m}^{(-v(k))}(\bx_k)$. The above variance estimator can be shown to be consistent, under mild conditions on the sampling design. We can thus justify an asymptotic $(1-\alpha)$ Wald-type confidence interval for
$\mu$,
\[
\widehat{\mu}_{\mathrm{MA}}^{\mathrm{cf}}(\widehat m)
\;\pm\;
z_{1-\alpha/2}\,
\left\{
\widehat \V_d\!\left(
\widehat{\mu}_{\mathrm{MA}}^{\mathrm{cf}}(\widehat m)
\right)
\right\}^{1/2}.
\]

\subsection{Cross-fitted model-assisted estimation under other sampling designs}\label{Cross_Other_sam}

\noindent In this section, we discuss the implementation of cross-fitting under more general sampling designs. For simplicity, we temporarily restrict attention to the case of two folds ($K=2$). The key simplification under Poisson sampling (see Section \eqref{Sec:POI_MA}) arises from the independence of the sampling indicators $(I_k)_{k\in U}$. By randomly partitioning $U$ into two folds $U_1$ and $U_2$, it follows that $(I_k)_{k\in U_1}$ and $(I_k)_{k\in U_2}$ are independent. Therefore, a regression function estimator $\widehat{m}_1$ constructed using observations from $U_1$ is independent of $(I_k)_{k\in U_2}$, leading to a particularly simple theoretical analysis. However, for most sampling designs, the sampling indicators $(I_k)_{k\in U_1}$ and $(I_k)_{k\in U_2}$ are generally not independent. In a causal inference setting, \cite{lu2025conditional} proposed a design-aware
cross-fitting construction in which the folds are built precisely to ensure
conditional independence across folds. Translated to survey sampling, their idea
can be used for SRSWOR, Bernoulli
sampling, and stratified SRSWOR. For instance, under SRSWOR with two folds, one splits the sampled units and the non-sampled units separately: the sample $S$ is partitioned into $S_1$ and $S_2$, and the non-sampled units $U\setminus S$ are partitioned into two corresponding
parts. The folds $U_1$ and $U_2$ are then formed by combining the sampled and
non-sampled units assigned to each fold. This construction is designed so that
\[
(I_k)_{k\in U_1}
\perp\!\!\!\perp
(I_k)_{k\in U_2}
\mid \boldsymbol{\delta},
\]
where $\boldsymbol{\delta}:=(\delta_{k,j}:k\in U,\;j=1,2)$ denotes the fold
indicators. This design-aware approach leads to particularly clean mathematical arguments
for studying the resulting estimators. \\

\noindent An alternative is to simply partition the population at random, as in the Poisson case. As noted above, independence between the fold indicators will typically not hold. However, another form of conditional independence arises in several designs; for example, it holds for some designs that
\begin{equation}\label{cond_ind}
	(I_k)_{k\in U_1} \perp\!\!\!\perp (I_k)_{k\in U_2} \mid \boldsymbol{\delta}, \boldsymbol{n},
\end{equation}
where $\boldsymbol{n} = (n_1, n_2)$ denotes the vector of sample sizes in each fold. This property arises naturally in several commonly used sampling designs, including SRSWOR, Poisson sampling, conditional Poisson sampling, and their stratifications. Informally, the dependence across folds is often driven by constraints on the sample sizes, such as $n = n_1 + n_2$. Conditionally on $\boldsymbol{n}$, this source of dependence is removed. These two crossfitting methodologies are closely related and can be used to create folds leading to valid inference. \\

\noindent Two variants of model-assisted estimators can be considered with cross-fitting: the estimator defined in \eqref{eq:MA_crossfit}, and a modified version given by
\begin{equation}\label{eq:MA_crossfit_modif}
	\widehat{\mu}_{\mathrm{MA}}^{\mathrm{cf}}(\widehat m, \tilde{\pi})
	=
	\frac{1}{N}\sum_{k\in U}
	\left\{
	\widehat m^{(-v(k))}(\mathbf{x}_k)
	+
	\frac{I_k}{\tilde{\pi}_k}
	\big(y_k-\widehat m^{(-v(k))}(\mathbf{x}_k)\big)
	\right\},
\end{equation}
where $\tilde{\pi}_k := \P (k \in S \mid \boldsymbol{\delta}, \boldsymbol{n})$ for $k \in U$.  In the case of Poisson sampling, the cross-fitted estimator $\widehat{\mu}_{\mathrm{MA}}^{\mathrm{cf}}(\widehat m)$ defined in \eqref{eq:MA_crossfit} is exactly unbiased due to the independence between folds. In contrast, under more general sampling designs, where only \eqref{cond_ind} holds, the estimator $\widehat{\mu}_{\mathrm{MA}}^{\mathrm{cf}}(\widehat m)$ is typically biased.  To see this, reproducing the calculations in \eqref{eq:cfUnbiased} yields, for $k \in U$,
\begin{align*}
	\E_d&\!\left[
	\widehat m^{(-v(k))}(\mathbf{x}_k)
	+
	\frac{I_k}{\pi_k}
	\big(y_k-\widehat m^{(-v(k))}(\mathbf{x}_k)\big)
	\;\middle|\;
	\{I_\ell\}_{\ell \in U_{-v(k)}}, \boldsymbol{\delta}, \boldsymbol{n}
	\right]\\&
	=
	\widehat m^{(-v(k))}(\mathbf{x}_k)
	+
	\frac{\tilde{\pi}_k}{\pi_k}
	\big(y_k-\widehat m^{(-v(k))}(\mathbf{x}_k)\big),
\end{align*}
which would be equal to $y_k$ if $\tilde{\pi}_k$ were used in place of the original first-order inclusion probability $\pi_k$. It follows that the estimator $\widehat{\mu}_{\mathrm{MA}}^{\mathrm{cf}}(\widehat m, \tilde{\pi})$ is design-unbiased for $\mu$. \\

\noindent Alternatively, it is possible to use  $\widehat{\mu}_{\mathrm{MA}}^{\mathrm{cf}}(\widehat m)$, at the price of a (typically small) bias. Under mild conditions, both estimators are asymptotically equivalent to their respective oracle counterparts, are root-$n$ consistent, and admit consistent variance estimators. Moreover, it can also be shown that, in some settings, the two estimators are first-order asymptotically equivalent, in the sense that
\[
\sqrt{n}\big(\widehat{\mu}_{\mathrm{MA}}^{\mathrm{cf}}(\widehat m)- \mu \big)
=
\sqrt{n}\big(\widehat{\mu}_{\mathrm{MA}}^{\mathrm{cf}}(\widehat m, \tilde{\pi})- \mu \big)
+ o_\P(1).
\]
Overall, as in the Poisson case, both estimators can be used to construct asymptotically valid confidence intervals. The reader is referred to \citet{an2026agnostic} for a more comprehensive discussion of the asymptotic properties of both model-assisted procedures.\\

\noindent  An important feature of the preceding results is the weakness of the conditions imposed on the fitted prediction rule. For first-order asymptotic validity, it is sufficient that the cross-fitted predictions be mean-square consistent for their population-level target $m_N$; no specific rate of convergence is required. This weak requirement results from two complementary features of the model-assisted construction. First, the corresponding estimating function is Neyman-orthogonal with respect to the outcome prediction rule. Second, the first-order inclusion probabilities $\pi_k$ are known under the sampling design and therefore introduce no additional estimation error. This favorable situation changes in the presence of missing outcomes. Because the response probabilities are generally unknown, they must themselves be estimated. The first-order remainder then depends jointly on the estimation errors of the outcome and response models, typically leading to a product-rate condition on the two nuisance estimators. We turn to this problem in the next section, in the context of imputation for survey nonresponse.

\section{IMPUTATION FOR THE TREATMENT OF ITEM NONRESPONSE}\label{Sec:IMP}
As mentioned in Section~\ref{Sec:INTRO}, we distinguish between unit
nonresponse and item nonresponse because they are typically handled using
different methods. Item nonresponse, the focus of this section, is usually
treated by single imputation, whereas unit nonresponse is commonly handled
through weight adjustment; the latter is discussed in Section~\ref{Sec:IPW}.
Table~\ref{Tab:nonreponses} illustrates this distinction. Respondents provide
usable values for all survey variables, units with item nonresponse provide
values for some but not all variables, and units with unit nonresponse provide
no survey information.\\
\begin{table}[h]
	\begin{center}
		\vspace{4mm}
		\begin{tabular}{c c c c c c c c }
			& $Y_1$ & $Y_2$ & $Y_3$ & \dots & $Y_p$  \\
			1 & \checkmark & \checkmark & \checkmark & \dots & \checkmark  & \ldelim \} {2}{1 mm} &  \multirow{2}{*}{\small{Respondents}}\\
			2 & \checkmark & \checkmark & \checkmark & \dots & \checkmark   \\
			\vdots & \checkmark &X  & X  & \dots  & \checkmark &  \ldelim \} {2}{1 mm} &  \multirow{2}{*}{\small{Item nonresponse}}\\
			\vdots & X & \checkmark & X & \dots & X  \\
			\vdots & X & X & X & \dots & X &  \ldelim \} {2}{1 mm} &  \multirow{2}{*}{\small{Unit nonresponse}}\\
			$n$ & X & X & X & \dots & X & 
		\end{tabular}
		
		\vspace{4mm}
		
		\caption{Unit vs. item nonresponse.}
		\label{Tab:nonreponses}
	\end{center}
\end{table}

\noindent Throughout this section, we assume the absence of unit
nonresponse, that is, all sampled units provide at least one observed
survey value. Moreover, item nonresponse is treated separately for each survey
variable. Accordingly, we focus on a generic survey variable $Y$, with the
understanding that imputation is applied independently
to each component $Y_1,\ldots,Y_p$. We again focus on estimating the finite population mean
$\mu = N^{-1}\sum_{k\in U} y_k$.\\

\noindent Let $r_k$ denote the potential item-response indicator for unit
$k\in U$, with $r_k=1$ if $y_k$ would be observed were unit $k$ selected
in the sample, and $r_k=0$ otherwise. Let
$\mathbf{r}=(r_1,\ldots,r_N)^\top$. The sets of sampled respondents and
sampled nonrespondents are
$S_r=\{k\in S:r_k=1\}$ and $S_m=\{k\in S:r_k=0\}$, respectively.
At the imputation stage, a vector of fully observed auxiliary variables
$\mathbf{x}_k$ is assumed to be available for all sampled units $k\in S$.
When design variables are related to the survey variable $Y$, they are
included among the auxiliary variables used in the imputation model.\\

\noindent In the remainder of this section, we work under the following assumptions.
First, we assume that the sampling design is non-informative for the
outcome model conditionally on the auxiliary variables, in the sense that
$$
\mathcal{F}(\mathbf{y}\mid \mathbf{I},\mathbf{X})
=
\mathcal{F}(\mathbf{y}\mid \mathbf{X}),
$$
where $\mathbf{y}=(y_1,\ldots,y_N)^\top$ and
$\mathbf{X}=(\mathbf{x}_1^\top,\ldots,\mathbf{x}_N^\top)^\top$. That is, after
conditioning on $\mathbf{X}$, sampling does not modify the relationship
between $\mathbf{y}$ and the auxiliary variables used in the imputation
model.\\

\noindent Second, we assume that the item-response mechanism is invariant to the
sampling design, conditional on $\mathbf{y}$ and $\mathbf{X}$, in the
sense that
$$
\mathcal{F}(\mathbf{r}\mid \mathbf{I},\mathbf{y},\mathbf{X})
=
\mathcal{F}(\mathbf{r}\mid \mathbf{y},\mathbf{X}).
$$
This condition allows the response indicators to be defined at the
finite-population level before the sample is selected. We further assume
that the item nonresponse mechanism is missing at random (MAR), that is,
$$
\mathds{P}(r_k=1 \mid y_k,\mathbf{x}_k)
=
\mathds{P}(r_k=1 \mid \mathbf{x}_k)
=: p(\mathbf{x}_k),
\qquad k\in U,
$$
where $p(\cdot)$ denotes the response propensity function. Together, the
response-invariance and MAR assumptions imply that
$$
\mathds{P}(r_k=1 \mid I_k=1,y_k,\mathbf{x}_k)
=
\mathds{P}(r_k=1 \mid \mathbf{x}_k)
=
p(\mathbf{x}_k).
$$
In addition, we assume that, conditionally on the auxiliary variables
$\mathbf{X}$, the response indicators $\{r_k:k\in U\}$ are independent
across units. We further assume a positivity condition: there exists a
constant $\delta>0$ such that
$$
p(\mathbf{x}_k)\ge \delta,
\qquad \text{a.s.}, \quad k\in U.
$$
The relationship between the survey variable $Y$ and $\mathbf{x}$ is
described by the imputation model
$$
y_k = m(\mathbf{x}_k) + \varepsilon_k,
\qquad k\in U,
$$
where $m(\mathbf{x})=\mathds{E}(y\mid\mathbf{x})$ is the unknown regression
function and the errors satisfy
$\mathds{E}(\varepsilon_k\mid\mathbf{x}_k)=0$ and
$\mathds{V}(\varepsilon_k\mid\mathbf{x}_k)=\sigma^2$, where $\sigma^2$ is an
unknown parameter.

\subsection{THE CUSTOMARY IMPUTED ESTIMATOR}\label{Sec:imp_est_cus}
If the conditional mean function $m(\cdot)$ were known, missing
values could be imputed by their oracle predictions $m(\mathbf{x}_k)$ for $k\in S_m$.
Define the oracle-imputed values as
\[
\widetilde{y}_{k}^{\rm OR}
=
r_k y_k + (1-r_k)m(\mathbf{x}_k),
\qquad k\in S,
\]
and consider the corresponding oracle imputed estimator of the finite
population mean $\mu=N^{-1}\sum_{k\in U}y_k$,
\begin{equation*}\label{eq:mu_or_imp}
	\widehat{\mu}_{I}(m)
	=
	\frac{1}{N}\sum_{k\in S} w_k\, 	\widetilde{y}_{k}^{\rm OR}.
\end{equation*}

\noindent Under the assumptions above, 
$\widehat{\mu}_{I}(m)$ is $mdq$-unbiased for $\mu$, in the
sense that
$
\E_m \E_d \E_q
\!\left\{
\widehat{\mu}_{I}(m)-\mu
\right\}
=
0.
$
Here, $\E_m(\cdot)$ denotes expectation with respect to the imputation model,
treating the finite population covariates $\mathbf{X}$, the sampling
indicators $\mathbf{I}$, and the response indicators $\mathbf{r}$ as fixed;
$\E_d(\cdot)$ denotes expectation with respect to the sampling design,
treating the finite population values $(\mathbf{y},\mathbf{X})$ and the
response indicators $\mathbf{r}$ as fixed; and $\E_q(\cdot)$ denotes
expectation with respect to the item nonresponse mechanism, treating the
finite population values $(\mathbf{y},\mathbf{X})$ and the sampling
indicators $\mathbf{I}$ as fixed.
Mdq-unbiasedness therefore means that the estimator is unbiased jointly
with respect to the superpopulation model for $Y$, the sampling design,
and the nonresponse mechanism.\\

\noindent Under standard regularity conditions, the oracle imputed estimator
$\widehat{\mu}_{I}(m)$ is root-$n$ consistent for $\mu$ and
satisfies a central limit theorem.
Specifically,
\[
\{	\V_{\mathrm{tot}}\!\left\{
\widehat{\mu}_{I}(m)\right\}\}^{-1/2}\,
\left\{
\widehat{\mu}_{I}(m)-\mu\right\}
\;\xrightarrow[n \to \infty]{\mathcal{L}}\;
\mathcal{N}\!\left(0,\,
1\right),
\]
where $	\V_{\mathrm{tot}}\!\left\{
\widehat{\mu}_{I}(m)\right\}$ denotes the total variance
of $\widehat{\mu}_{I}(m)$. Adopting the so-called reverse approach originally proposed by \cite{fay1991design} and further developed by \cite{shao1999variance}, the variance of $\widehat{\mu}_{I}(m)$ can be expressed as
\begin{align}\label{Vtot_OR_imp}
	\V_{\mathrm{tot}}\!\left\{
	\widehat{\mu}_{I}(m)
	\right\}
	&=
	\E_m\E_q\V_d\!\left\{
	\widehat{\mu}_{I}(m)
	\right\}
	+
	\E_q
	\V_m\!
	\E_d\!\left\{
	\widehat{\mu}_{I}(m)-\mu
	\right\}.
\end{align}
\noindent Although the oracle estimator is not feasible in practice, it nevertheless provides a useful benchmark, as it is unbiased, root-$n$ consistent, and asymptotically normal under the $mdq$ framework.\\

\noindent In practice, $m(\cdot)$ is unknown and must be estimated from the respondents. 
Let $\widehat m(\cdot)$ denote an estimator of $m(\cdot)$ constructed using 
$\{(y_k,\mathbf{x}_k):k\in S_r\}$.  Here, $m(\cdot)$ plays the role of a nuisance function: it is not itself
of primary interest, but serves as an ingredient in the construction of an
estimator of the population mean $\mu$. We define the feasible imputed values
\[
\widetilde{y}_{k} = r_k y_k + (1-r_k)\widehat{m}(\mathbf{x}_k), 
\qquad k\in S.
\]
The feasible imputed estimator of $\mu$ is then
\begin{equation*}\label{eq:mu_feas_imp}
	\widehat{\mu}_{I}(\widehat{m}) 
	= \frac{1}{N}\sum_{k\in S} w_k\, \widetilde{y}_{k}.
\end{equation*}
This estimator can be expressed as
\begin{equation}\label{eq:imp_decomp}
	\widehat{\mu}_{I}(\widehat{m})
	=
	\widehat{\mu}_{I}(m)
	+
	\frac{1}{N}\sum_{k\in S} w_k(1-r_k)
	\Big\{\widehat m(\mathbf{x}_k)-m(\mathbf{x}_k)\Big\}.
\end{equation}
The feasible estimator $\widehat{\mu}_{I}(\widehat m)$ is not Neyman orthogonal with respect to the imputation model $m(\cdot)$; see Section S1 of the Supplementary Material. Indeed, the second term on the right-hand side of \eqref{eq:imp_decomp} captures the effect of estimating $m(\cdot)$ for the nonresponding units. In this sense, it represents the price paid for replacing the unobserved values by predictions. Since this term enters linearly in $\widehat m(\cdot)-m(\cdot)$, errors in estimating $m(\cdot)$ may affect $\widehat{\mu}_{I}(\widehat m)$ at the first order. The implications of this lack of orthogonality depend on how $m(\cdot)$ is estimated and are discussed next.\\

\noindent 
To better understand the behavior of the remainder term in
\eqref{eq:imp_decomp}, it is useful to first consider the case of
deterministic linear regression imputation, which belongs to the class of
parametric imputation procedures.
Suppose that
$m(\mathbf{x}_k)=\mathbf{x}_k^\top\boldsymbol{\beta}$,
where $\boldsymbol{\beta}$ is estimated from the respondents using the
least squares estimator $\widehat{\boldsymbol{\beta}}$.
Then, the remainder term in \eqref{eq:imp_decomp} becomes
$
N^{-1}\sum_{k\in S} w_k(1-r_k)\mathbf{x}_k^{\top}(\widehat{\boldsymbol{\beta}}-\boldsymbol{\beta}).
$
\noindent
Under standard regularity conditions,
$\widehat{\boldsymbol{\beta}}-\boldsymbol{\beta}=O_{\P}(n^{-1/2})$.
Consequently, the remainder term is typically of order $O_{\P}(n^{-1/2})$.
The feasible imputed estimator $\widehat{\mu}_{I}(\widehat{m})$ is therefore
root-$n$ consistent, but it is not generally asymptotically equivalent to the
oracle estimator $\widehat{\mu}_{I}(m)$. Because the estimation error in $\widehat m(\cdot)$ enters the estimator at
the first order, the asymptotic variance of
$\widehat{\mu}_{I}(\widehat m)$ contains, in addition to the oracle variance
in \eqref{Vtot_OR_imp}, a component reflecting the variability introduced by
estimating $\boldsymbol{\beta}$. This additional variance term arises directly from the lack of Neyman
orthogonality of the customary imputed estimator 	$\widehat{\mu}_{I}(\widehat{m})$.
In parametric settings, however, this extra variability can be handled using
standard linearization techniques, and a substantial literature exists on
variance estimation accounting for sampling and nonresponse; see, for example, \cite{sarndal1992methods, rao1992jackknife, shao1996bootstrap, shao1999variance, kim2009unified, haziza2020variance}.
Central limit theorems for imputed estimators in such settings are discussed,
for instance, in \cite{an2026variable}.\\

\noindent The situation becomes more complicated when $m(\cdot)$ is estimated using
flexible nonparametric or ML methods. Although these methods may improve
predictive accuracy, their estimation error often converges more slowly than
$1/\sqrt{n}$. Because the customary imputed estimator is not Neyman
orthogonal, this error enters directly through the remainder term in
\eqref{eq:imp_decomp} If the prediction error
$\widehat m(\mathbf{x}_k)-m(\mathbf{x}_k)$ does not decrease sufficiently
fast, the remainder may dominate the sampling variability. The estimator may
still be consistent for $\mu$ if prediction errors vanish on average, but
root-$n$ consistency and a central limit theorem need not hold. In such cases,
the additional variability induced by estimating $m(\cdot)$ is difficult to
characterize, making reliable variance estimation challenging.\\

\noindent
From a practical perspective, this means that ML-based single imputation may
still perform well for point estimation in large samples, but it generally
does not support valid statistical inference without modifying the
estimation strategy.
This observation motivates the use of the double/debiased ML (DML)
framework of Chernozhukov et al.\ (2018), which combines Neyman orthogonality
with cross-fitting to control the impact of nuisance estimation and can restore valid root-$n$ inference under suitable conditions. These ideas are developed in the next section.

\subsection{THE ORACLE AIPW ESTIMATOR}
\noindent
We begin by introducing an oracle version of the augmented inverse
probability weighted (AIPW) estimator for the finite population mean
$\mu$. This estimator assumes that both the outcome regression function
$m(\cdot)$ and the response propensity score
$p(\cdot)$ are known. Although unrealistic in practice, this
assumption provides a useful benchmark for developing feasible
estimators. The oracle AIPW estimator of $\mu$ is defined as
\begin{align*}\label{DR_OR}
	\widehat{\mu}_{\mathrm{aipw}}(m,p)
	=
	\frac{1}{N}
	\left\{
	\sum_{k \in S} w_k\, m(\mathbf{x}_k)
	+
	\sum_{k \in S_r} w_k\,
	\frac{y_k - m(\mathbf{x}_k)}{p(\mathbf{x}_k)}
	\right\}
	\equiv
	\frac{1}{N}\sum_{k \in S} w_k\, \eta_k,
\end{align*}
where the pseudo-values $\eta_k$ are given by
\begin{equation*}\label{pseudo-value_DR_OR}
	\eta_k
	=
	m(\mathbf{x}_k)
	+
	\frac{r_k}{p(\mathbf{x}_k)}
	\big\{y_k - m(\mathbf{x}_k)\big\},
	\qquad k \in U.
\end{equation*}
The estimator $\widehat{\mu}_{\mathrm{aipw}}(m,p)$ can thus be viewed as
a Horvitz--Thompson type estimator applied to pseudo-values. The estimator $\widehat{\mu}_{\mathrm{aipw}}(m,p)$ is unbiased for $\mu$,
in the sense that
\[
\E_m \E_d \E_q\!\left\{
\widehat{\mu}_{\mathrm{aipw}}(m,p)-\mu
\right\}
=0.
\]
Moreover,
under mild regularity conditions,
$\widehat{\mu}_{\mathrm{aipw}}(m,p)$ is root-$n$ consistent for $\mu$;
see \cite{dagdoug2026debiased}.\\

\noindent
To construct variance estimators for $\widehat{\mu}_{\mathrm{aipw}}(m,p)$,
we again adopt the reverse approach (see Section
\eqref{Sec:imp_est_cus}), whereby the total variance of
$\widehat{\mu}_{\mathrm{aipw}}(m,p)$ is given by
\begin{align}\label{Vtot_OR}
	\V_{\mathrm{tot}}\!\left\{
	\widehat{\mu}_{\mathrm{aipw}}(m,p)
	\right\}
	&=
	\E_m\E_q\V_d\!\left\{
	\widehat{\mu}_{\mathrm{aipw}}(m,p)
	\right\}
	+
	\E_q
	\V_m
	\E_d\!\left\{
	\widehat{\mu}_{\mathrm{aipw}}(m,p)-\mu
	\right\}\nonumber\\
	&=
	\E_m\E_q\left(\frac{1}{N^2}
	\sum_{k \in U}\sum_{\ell \in U}
	\Delta_{k\ell}\,
	\frac{\eta_k}{\pi_k}\,
	\frac{\eta_\ell}{\pi_\ell}\right)
	+
	\frac{\sigma^2}{N^2}
	\sum_{k \in U}
	\frac{1-p(\mathbf{x}_k)}{p(\mathbf{x}_k)} \nonumber\\
	&\equiv
	\V_1\!\left\{
	\widehat{\mu}_{\mathrm{aipw}}(m,p)
	\right\}
	+
	\V_2\!\left\{
	\widehat{\mu}_{\mathrm{aipw}}(m,p)
	\right\}.
\end{align}
The first component
$\V_1\!\{\widehat{\mu}_{\mathrm{aipw}}(m,p)\}$ corresponds to the design
variance of a Horvitz--Thompson–type estimator (see
Equation~\eqref{eq:Var_HT_mean}) in which the survey variable $y_k$ is
replaced by the pseudo-value $\eta_k$. Consequently, $\V_1\!\{\widehat{\mu}_{\mathrm{aipw}}(m,p)\}$ can be
estimated using any complete-data variance estimation procedure
available for Horvitz--Thompson estimators. Under mild regularity conditions, the contribution of the second component
$\V_2\!\{\widehat{\mu}_{\mathrm{aipw}}(m,p)\}$ is negligible when the
sampling fraction $n/N$ is small \citep{shao1999variance}.\\

\noindent Assuming that the conditional variance $\sigma^2$ is known,
a $mdq$-unbiased estimator of the total variance in \eqref{Vtot_OR} can be
obtained by estimating each component without bias, leading to
\begin{equation}\label{Vtot_OR_hat}
	\widehat{\V}_{\mathrm{tot}}\!\left\{
	\widehat{\mu}_{\mathrm{aipw}}(m,p)
	\right\}
	=
	\frac{1}{N^2}
	\sum_{k \in S}\sum_{\ell \in S}
	\frac{\Delta_{k\ell}}{\pi_{k\ell}}\,
	\frac{\eta_k}{\pi_k}\,
	\frac{\eta_\ell}{\pi_\ell}
	+
	\frac{	\sigma^2}{N^2}
	\sum_{k \in S_r}
	w_k\,
	\frac{1-p(\mathbf{x}_k)}{\{p(\mathbf{x}_k)\}^2}.
\end{equation}
That is,
$$\E_m\E_d\E_q \left[	\widehat{\V}_{\mathrm{tot}}\!\left\{
\widehat{\mu}_{\mathrm{aipw}}(m,p)
\right\}\right]=	\V_{\mathrm{tot}}\!\left\{
\widehat{\mu}_{\mathrm{aipw}}(m,p)
\right\}.$$
In Section~\eqref{Sec:AIPW:feas}, for simplicity, we assume that the sampling
fraction $n/N$ is negligible, so that the second term in
\eqref{Vtot_OR_hat} can be omitted. Estimation of this term is discussed in
\cite{dagdoug2026debiased}. Finally, under appropriate regularity conditions
\citep{dagdoug2026debiased}, the oracle AIPW estimator satisfies a central
limit theorem:
\[
\left\{
\widehat{\V}_{\mathrm{tot}}\!\left(
\widehat{\mu}_{\mathrm{aipw}}(m,p)
\right)
\right\}^{-1/2}
\big(
\widehat{\mu}_{\mathrm{aipw}}(m,p)-\mu
\big)
\;\xrightarrow[n\to \infty]{\mathcal{L}}\;
\mathcal{N}(0,1).
\]

\subsection{THE FEASIBLE CROSS-FITTED AIPW ESTIMATOR}\label{Sec:AIPW:feas}
In practice, the outcome regression function $m(\cdot)$ and the response
propensity function $p(\cdot)$ are unknown and must be estimated from the
observed data. Let $\widehat m(\cdot)$ and $\widehat p(\cdot)$ denote
estimators of $m(\cdot)$ and $p(\cdot)$, respectively, constructed using the
sample information. Although $m(\cdot)$ and $p(\cdot)$ must be estimated in practice,
they are not parameters of primary interest. Their role is simply to
help construct an estimator of the population mean $\mu$. The feasible AIPW estimator of $\mu$ is 
\begin{equation*}\label{eq:AIPW_feasible}
	\widehat{\mu}_{\mathrm{aipw}}(\widehat m,\widehat p)
	=
	\frac{1}{N}
	\sum_{k\in S}
	w_k\widehat{\eta}_k,
\end{equation*}
where 
\begin{equation}\label{pseudo-value_DR_hat}
	\widehat{\eta}_k
	=
	\widehat{m}(\mathbf{x}_k)
	+
	\frac{r_k}{\widehat{p}(\mathbf{x}_k)}
	\big\{y_k - \widehat{m}(\mathbf{x}_k)\big\},
	\qquad k \in S.
\end{equation}
To assess the impact of estimating the nuisance functions with $\widehat{m}(\cdot)$ and  $\widehat{p}(\cdot)$, it is useful to
express the feasible AIPW estimator as
\begin{align*}\label{eq:AIPW_decomp_slide}
	\widehat{\mu}_{\mathrm{aipw}}(\widehat m,\widehat p)
	&=
	\widehat{\mu}_{\mathrm{aipw}}(m,  p)
	+
	R_n(\widehat m,\widehat p),
\end{align*}
where the remainder term can be written as
\[
R_n(\widehat m,\widehat p)=T_{1n}+T_{2n}+T_{3n},
\]
with
\begin{align*}
	T_{1n}
	&=
	\frac{1}{N}\sum_{k\in S} w_k
	\left\{1-\frac{r_k}{p(\mathbf{x}_k)}\right\}
	\big\{\widehat{m}(\mathbf{x}_k)-m(\mathbf{x}_k)\big\},\\
	T_{2n}
	&=
	-\frac{1}{N}\sum_{k\in S} w_k\,
	\frac{r_k}{p(\mathbf{x}_k)}	\frac{1}{\widehat{p}(\mathbf{x}_k)}\big\{y_k-{m}(\mathbf{x}_k)\big\}
	\left\{
	\widehat{p}(\mathbf{x}_k)-	{p(\mathbf{x}_k)}	
	\right\},\\
	T_{3n}
	&=
	\frac{1}{N}\sum_{k\in S} w_k\,
	\frac{r_k}{p(\mathbf{x}_k)}\frac{1}{\widehat{p}(\mathbf{x}_k)}
	\big\{\widehat{m}(\mathbf{x}_k)-m(\mathbf{x}_k)\big\}
	\left\{
	\widehat{p}(\mathbf{x}_k)-	{p(\mathbf{x}_k)}
	\right\}.
\end{align*}

\noindent A natural question is whether the remainder term
$R_n(\widehat m,\widehat p)$ is asymptotically negligible.
If it is, then the feasible estimator
$\widehat{\mu}_{\mathrm{aipw}}(\widehat m,\widehat p)$ has the same
first-order asymptotic behavior as the oracle estimator
$\widehat{\mu}_{\mathrm{aipw}}(m,p)$.
This is typically what happens when the nuisance functions are estimated
by correctly specified parametric models: the estimation errors in
$\widehat m(\cdot)$ and $\widehat p(\cdot)$ are then small enough for
the remainder term to be of smaller order than $1/\sqrt{n}$, so that the
feasible and oracle estimators are asymptotically equivalent.\\

\noindent The situation is more delicate when $m(\cdot)$ and $p(\cdot)$ are
estimated using flexible ML methods. Although the score function underlying the AIPW estimator satisfies Neyman orthogonality  (see Section S1 of the Supplementary Material) this
property alone does not guarantee that the remainder term
$R_n(\widehat m,\widehat p)$ is asymptotically negligible.
The issue arises because the same realized sample is used twice: first
to estimate the nuisance functions, and then again to construct the
AIPW estimator.
Because of this reuse of the data, the estimation errors in
$\widehat m(\cdot)$ and $\widehat p(\cdot)$ may remain statistically
dependent on the sampling indicators $I_k$, the response indicators $r_k$, and on the observed outcomes $y_k$
appearing in the estimator.
This is precisely why the asymptotic negligibility of
$R_n(\widehat m,\widehat p)$ is no longer automatic in flexible
nonparametric settings.\\

\noindent The decomposition above separates first-order and second-order effects.
The terms $T_{1n}$ and $T_{2n}$ are first-order terms: $T_{1n}$ is linear
in $\widehat m-m$, whereas $T_{2n}$ is linear in $\widehat p-p$. By contrast,
$T_{3n}$ is a second-order interaction term, since it involves the product of
the two nuisance estimation errors. This raises the question of whether the
first-order terms can be made negligible, so that the remainder is driven by
the interaction term.\\

\noindent Without cross-fitting, and without strong assumptions on the underlying function class, this is generally not the case. When the same data are used both to estimate the nuisance functions and to evaluate the estimating equation, the quantities $\widehat m(\mathbf{x}_k)-m(\mathbf{x}_k)$ and $\widehat p(\mathbf{x}_k)-p(\mathbf{x}_k)$ may remain tied to the same observations that generate $r_k$ and $y_k$. In particular, the summands in $T_{1n}$ and $T_{2n}$ are not necessarily centered, in the sense that their conditional expectations, given the fitted nuisance functions, need not be zero. For example, without sample splitting, the factor $\widehat m(\mathbf{x}_k)-m(\mathbf{x}_k)$ may still depend on the same response indicator $r_k$ that appears in $1-r_k/p(\mathbf{x}_k)$; similarly, $\widehat p(\mathbf{x}_k)-p(\mathbf{x}_k)$ may still depend on the same observed outcome contribution $r_k\{y_k-m(\mathbf{x}_k)\}$. As a result, even though the AIPW score is Neyman orthogonal at the population level, the empirical first-order terms $T_{1n}$ and $T_{2n}$ may still contribute at the $n^{-1/2}$ scale when the nuisance functions are trained and evaluated on the same data.\\

\noindent Cross-fitting, described below, is introduced precisely to break this dependence.
The sample is partitioned into several folds, and each unit is evaluated
using nuisance estimators trained on the other folds only.
Once this is done, the first-order terms become conditionally centered,
at least asymptotically, because the estimation errors are no longer
tied to the same observations that enter the estimating equation.
What remains is the second-order term $T_{3n}$. That is, after cross-fitting, the behavior of the remainder
is driven by the product of the two nuisance estimation errors.\\

\noindent This is why the relevant conditions involve both consistency of the nuisance estimators and a product-rate condition. We assume that
\begin{equation}\label{nuisance_consistency}
	\|\widehat m-m\|=o_{\P}(1),
	\qquad
	\|\widehat p-p\|=o_{\P}(1),
\end{equation}
and that
\begin{equation}\label{prod_rate}
	\|\widehat m-m\|\times \|\widehat p-p\| = o_{\P}(n^{-1/2}),
\end{equation}
where $\|\cdot\|$ denotes an appropriate $L_2$ norm. Together with cross-fitting and standard regularity conditions, the consistency conditions in \eqref{nuisance_consistency} make the first-order terms $T_{1n}$ and $T_{2n}$ negligible, while the product-rate condition in \eqref{prod_rate} controls the second-order interaction term $T_{3n}$.
Under these conditions,
\[
R_n(\widehat m,\widehat p)=o_{\P}(n^{-1/2}).
\]
Therefore, the feasible estimator is asymptotically equivalent to the oracle estimator and inherits its first-order asymptotic properties, including root-$n$ consistency and asymptotic normality.\\

\noindent It is worth emphasizing that Condition \eqref{prod_rate} does not require
either nuisance estimator to converge at the parametric rate
$n^{-1/2}$.
What matters is the product of the two errors.
For example, if
$\|\widehat m-m\|=O_\P(n^{-1/4})$
and
$\|\widehat p-p\|=O_\P(n^{-1/3})$,
then
\[
\|\widehat m-m\|\ \times \|\widehat p-p\|=O_\P(n^{-7/12})
=o_\P(n^{-1/2}),
\]
so that \eqref{prod_rate} is satisfied.
More generally, if the two convergence exponents add up to a value
strictly larger than $1/2$, then the product-rate condition holds.
This is one of the main practical advantages of combining Neyman-orthogonal score functions with cross-fitting: valid root-$n$ inference
remains possible even when each nuisance estimator converges relatively
slowly on its own.\\

\noindent The same dependence issue also affects variance estimation.
Assuming that the sampling fraction $n/N$ is negligible, one might be
tempted to estimate the variance by replacing $\eta_k$ in the first term
on the right-hand side of \eqref{Vtot_OR_hat} with the plug-in values
$\widehat{\eta}_k$ defined in \eqref{pseudo-value_DR_hat}
However, these pseudo-values depend directly on the estimated residuals
$y_k-\widehat m(\mathbf{x}_k)$ and on the estimated propensities
$\widehat p(\mathbf{x}_k)$.
When highly adaptive ML methods are used, overfitting may
make the residuals artificially small, which can lead to a
serious downward bias in plug-in variance estimators.\\

\noindent To address these dependence and overfitting issues, a standard approach is
to use cross-fitting. Fix an integer $K\ge 2$ and partition the realized
sample $S$ into $K$ disjoint folds $S_1,\ldots,S_K$. For a sampled unit
$k\in S$, let $v(k)\in\{1,\ldots,K\}$ denote the index of its fold, and write
$S_{-v}=S\setminus S_v$. For each fold $v$, estimate $m(\cdot)$ using the
respondents in $S_{-v}$ and estimate $p(\cdot)$ using the response indicators
in $S_{-v}$, yielding $\widehat m^{(-v)}(\cdot)$ and
$\widehat p^{(-v)}(\cdot)$. The same fold partition is used for both nuisance
functions, and unit $k$ is evaluated using
$\widehat m^{(-v(k))}(\mathbf{x}_k)$ and
$\widehat p^{(-v(k))}(\mathbf{x}_k)$.\\

\noindent
The cross-fitted feasible AIPW estimator can then be written as
\begin{equation*}\label{eq:AIPW_cf}
	\widehat{\mu}_{\mathrm{aipw}}^{\mathrm{cf}}(\widehat m,\widehat p)
	=
	\frac{1}{N}
	\sum_{k\in S}
	w_k \widehat{\eta}_k^{\mathrm{cf}},
\end{equation*}
where
\[
\widehat{\eta}_k^{\mathrm{cf}}
=
\widehat m^{(-v(k))}(\mathbf{x}_k)
+
\frac{r_k}{\widehat p^{(-v(k))}(\mathbf{x}_k)}
\big\{y_k-\widehat m^{(-v(k))}(\mathbf{x}_k)\big\}.
\]
Here, for each sampled unit $k$, the nuisance estimators
$\widehat m^{(-v(k))}(\cdot)$ and $\widehat p^{(-v(k))}(\cdot)$ are trained
using only observations in the complementary subsample
$S_{-v(k)}=S\setminus S_{v(k)}$.\\

\noindent Under mild regularity conditions, the cross-fitted estimator satisfies
\[
\sqrt{n}\Big(
\widehat{\mu}_{\mathrm{aipw}}^{\mathrm{cf}}(\widehat m,\widehat p)
-
\widehat{\mu}_{\mathrm{aipw}}(m,p)
\Big)
=
o_{\P}(1),
\]
provided that Conditions \eqref{nuisance_consistency} and \eqref{prod_rate} hold. When this happens,
$\widehat{\mu}_{\mathrm{aipw}}^{\mathrm{cf}}(\widehat m,\widehat p)$
is asymptotically equivalent to the oracle estimator
$\widehat{\mu}_{\mathrm{aipw}}(m,p)$ and therefore inherits its
first-order asymptotic properties. In practice, however, the convergence rates of the nuisance estimators are
rarely known. For many nonparametric methods, they depend on the smoothness
of the unknown function and on the dimension of the covariates
\citep{stone1982optimal}. It is therefore often unclear whether Condition
\eqref{prod_rate} holds in a given application. One practical response is to
use aggregation methods, such as ensemble or Super Learner approaches
\citep{van2007super}. Separate libraries may be used for the outcome
regression $m(\cdot)$ and the response propensity $p(\cdot)$; by combining
estimators with different assumptions, the aggregated estimators may adapt to
the best-performing method in each library, increasing the chances that
Condition \eqref{prod_rate} holds.\\

\noindent Finally, assuming that the sampling fraction $n/N$ is negligible, a consistent variance estimator (Dagdoug and Haziza, 2026) is obtained by replacing $\eta_k$ in the first term on the right-hand side of \eqref{Vtot_OR_hat} with $\widehat{\eta}_k^{\mathrm{cf}},$ leading to
\begin{equation*}\label{Vtot_OR_hat_cf}
	\widehat{\V}_{\mathrm{tot}}\!\left\{
	\widehat{\mu}_{\mathrm{aipw}}^{\mathrm{cf}}(\widehat m,\widehat p)
	\right\}
	=
	\frac{1}{N^2}
	\sum_{k \in S}\sum_{\ell \in S}
	\frac{\Delta_{k\ell}}{\pi_{k\ell}}\,
	\frac{\widehat{\eta}_k^{\mathrm{cf}}}{\pi_k}\,
	\frac{\widehat{\eta}_\ell^{\mathrm{cf}}}{\pi_\ell}.
\end{equation*}
Consequently, an asymptotic $(1-\alpha)$ Wald-type confidence interval for $\mu$
is given by
\begin{equation*}\label{eq:Wald_CI_cf_AIPW}
	\widehat{\mu}_{\mathrm{aipw}}^{\mathrm{cf}}(\widehat m,\widehat p)
	\ \pm\
	z_{1-\alpha/2}\,
	\left\{
	\widehat{\V}_{\mathrm{tot}}\!\left(
	\widehat{\mu}_{\mathrm{aipw}}^{\mathrm{cf}}(\widehat m,\widehat p)
	\right)
	\right\}^{1/2}.
\end{equation*}

\noindent
We end this section with two important remarks.

\medskip
\noindent
\textbf{Remark 1.}
The success of the DML approach depends crucially on the ability of the
ML methods to approximate the nuisance functions $m(\cdot)$ and $p(\cdot)$
well. From a theoretical perspective, the validity of the cross-fitted
AIPW estimator requires both consistency of the nuisance estimators, as in
\eqref{nuisance_consistency}, and a sufficiently fast product rate, as in
\eqref{prod_rate}. Thus, the methods used to construct $\widehat m$ and
$\widehat p$ must be able to estimate the outcome regression and response
propensity functions with enough accuracy for these two conditions to hold. In practice, the performance of the procedure therefore depends on the
choice of the architecture.
In this sense, there is no ``free lunch'': if the methods used to estimate
$m$ and $p$ are poorly chosen, one or both nuisance estimators may fail to
converge to their targets, and the product-rate condition may also fail.
The resulting DML estimator may then lead to unreliable inference.
Cross-validation can be used to guide the selection of hyperparameters and
improve predictive performance. However, the choice and tuning of ML
architectures in the context of DML for survey nonresponse remain an active
area of research.

\medskip
\noindent
\textbf{Remark 2.}
From an operational perspective, the cross-fitted AIPW estimator $\widehat{\mu}_{\mathrm{aipw}}^{\mathrm{cf}}(\widehat m,\widehat p)$ is not expressed in a form that is immediately convenient for implementation. In practice, national statistical offices typically disseminate completed microdata files in which each survey variable is represented by a single column containing observed values for respondents and imputed values for nonrespondents, together with a single column of survey weights. Secondary users are then accustomed to computing estimates of population totals or means using simple weighted sums. By contrast, the cross-fitted AIPW estimator does not naturally admit a representation as a weighted sum of observed and imputed values using a set of weights. To overcome this issue, we propose the following implementation strategy: For responding units ($r_k=1$), we report the observed values $y_k$ in the released data file. For nonresponding units ($r_k=0$), we report modified imputed values defined by 
\begin{equation}\label{mod_imp_val} \widehat{m}^{*}(\mathbf{x}_k) := \widehat m^{(-v(k))}(\mathbf{x}_k) + \frac{1}{\sum_{\ell\in S_m} w_\ell} \sum_{\ell\in S_r} w_\ell \left\{ \frac{1-\widehat p^{(-v(\ell))}(\mathbf{x}_\ell)} {\widehat p^{(-v(\ell))}(\mathbf{x}_\ell)} \right\} \left\{ y_\ell-\widehat m^{(-v(\ell))}(\mathbf{x}_\ell) \right\}. 
\end{equation} With the modified imputed values \eqref{mod_imp_val}, it is straightforward to verify that \begin{equation*}\label{cal_con} 
	\frac{1}{N} \sum_{k \in S} w_k \left\{ r_k y_k + (1-r_k)\widehat{m}^{*}(\mathbf{x}_k) \right\} = \widehat{\mu}_{\mathrm{aipw}}^{\mathrm{cf}}(\widehat m,\widehat p). 
\end{equation*} Consequently, the cross-fitted AIPW estimator can be computed using customary point-estimation procedures applied to a single completed data file with standard survey weights.

\section{INVERSE PROBABILITY WEIGHTING FOR THE TREATMENT OF UNIT NONRESPONSE}\label{Sec:IPW}

Unit nonresponse in surveys (see Table~\ref{Tab:nonreponses}) is typically
handled through weight adjustment. Response propensities are estimated using
auxiliary information available for both respondents and nonrespondents, and
the initial design weights $w_k=1/\pi_k$ are modified to produce a single set
of nonresponse-adjusted weights. These weights are then applied uniformly to
all survey variables $Y_1,\ldots,Y_p$. This variable-agnostic approach is a
defining feature of survey weighting: it preserves coherence across levels of
aggregation and maintains known relationships among survey variables at the
estimation stage. See, for example, \cite{haziza2017construction}.\\

\noindent 
In this section, we focus exclusively on unit nonresponse and ignore item
nonresponse. Thus, for a given sampled unit, either all survey variables
$Y_1,\ldots,Y_p$ are observed (respondents) or none of them is observed
(nonrespondents). In what follows, we consider a generic survey variable
$y$ and focus on estimating its population mean
$\mu = N^{-1}\sum_{k \in U} y_k$.
Let $r_k$ denote the unit response indicator, where $r_k=1$ if unit $k$
responds to the survey and $r_k=0$ otherwise. We assume that unit nonresponse is missing at random (MAR), that is,
$\Pr(r_k=1\mid y_k,\mathbf{x}_k)=\Pr(r_k=1\mid \mathbf{x}_k)=p(\mathbf{x}_k)$.  In addition, we assume that, conditional on the auxiliary variables, the response
indicators $\{r_k : k\in U\}$ are independent across units, with
\[
r_k \mid \mathbf{x}_k \sim \text{Bernoulli}\{p(\mathbf{x}_k)\},
\qquad k\in U.
\]	
Finally, for the theoretical results, we assume a positivity condition:
there exists a constant $\delta>0$ such that
$p(\mathbf{x}_k)\ge \delta$ for all $k\in U$. This condition rules out
arbitrarily small response probabilities and ensures that the inverse
probability weights remain stable. In practice, positivity is an
idealization, as most surveys contain a small fraction of so-called
hardcore nonrespondents whose probability of response is effectively zero.

\subsection{THE ORACLE IPW ESTIMATOR}
Assume for now that $p(\mathbf{x}_k)$ is known for all $k\in U$. The oracle
inverse probability weighted (IPW) estimator of the population mean $\mu$ is
\begin{equation}\label{eq:oracle_IPW_mean}
	\widehat{\mu}_{\mathrm{ipw}}(p)
	=
	\frac{1}{N}\sum_{k\in S_r}
	\,\frac{w_k}{p(\mathbf{x}_k)}\,y_k
\end{equation}
where $S_r=\{k\in S: r_k=1\}$ denotes the set of responding units. Estimator~\eqref{eq:oracle_IPW_mean} is the classical double expansion
estimator encountered in two-phase sampling (e.g., \cite{sarndal1992model}
Chapter~9): the first phase corresponds to the sample selection (with expansion
weight $w_k=1/\pi_k$), while unit nonresponse can be viewed as a second-phase
selection mechanism acting on $S$, with second-phase expansion factor
$1/p(\mathbf{x}_k)$.\\

\noindent
In imputation (see Section~\eqref{Sec:IMP}), inference is conducted for a single
survey variable $y$, and the imputation model is specified directly for that
variable. As a result, the $mdq$ inferential framework naturally accounts for
uncertainty arising from the superpopulation model, the sampling design, and
the nonresponse mechanism. By contrast, the treatment of unit nonresponse relies
on the construction of a single set of response-adjusted weights that is
applied uniformly to all survey variables $Y_1,\ldots,Y_p$. Because these weights are not tailored to a specific survey variable, inference for IPW-based estimators is more naturally conducted under a $dq$
framework, where the finite-population vectors $\mathbf{y}$ and
$\mathbf{X}$ are treated as fixed and the properties of the
estimators are derived with respect to the joint distribution induced by the
sampling design and the nonresponse mechanism. Since $\E_q(r_k\mid \mathbf{x}_k)=p(\mathbf{x}_k)$ and $\E_d(I_k w_k)=1$,
we have $\E_d\E_q\{\widehat{\mu}_{\mathrm{ipw}}(p)\}=\mu$.
The total $dq$-variance of 
$\widehat{\mu}_{\mathrm{ipw}}(p)$ can be expressed as
\begin{align}\label{eq:Vtot_IPW_twophase}
	\V_{dq}\!\left\{\widehat{\mu}_{\mathrm{ipw}}(p)\right\}
	&=
	\V_d
	\E_q\!\left\{
	\widehat{\mu}_{\mathrm{ipw}}(p)
	\right\}
	+
	\E_d
	\V_q\!\left\{
	\widehat{\mu}_{\mathrm{ipw}}(p)
	\right\}\nonumber\\
	&=
	\V_d\!\left(\widehat{\mu}_{\mathrm{HT}}\right)
	+
	\frac{1}{N^2}
	\sum_{k\in U}
	w_k\,
	\frac{1-p(\mathbf{x}_k)}{p(\mathbf{x}_k)}\,
	y_k^2,
\end{align}
where the first term corresponds to the sampling variance of the full-sample
estimator $\widehat{\mu}_{\mathrm{HT}}$, while the second term represents the additional variability
introduced by unit nonresponse. Expression~\eqref{eq:Vtot_IPW_twophase} also
shows that the contribution of nonresponse to the variance depends on the
inverse of the response probabilities. In particular, small values of
$p(\mathbf{x}_k)$ inflate the variance contributions, reflecting the familiar
instability of IPW estimators when the positivity condition is weak.\\

\noindent
An estimator of the total variance in
\eqref{eq:Vtot_IPW_twophase} can be obtained by  estimating the sampling and
nonresponse components separately, leading to
\begin{align}
	\widehat{\V}\!\left\{\widehat{\mu}_{\mathrm{ipw}}(p)\right\}
	&=
	\frac{1}{N^2}
	\left\{
	\sum_{k\in S_r}\sum_{\ell\in S_r}
	\frac{\Delta_{k\ell}}{\pi_{k\ell}}\,
	\frac{y_k}{\pi_k p(\mathbf{x}_k)}\,
	\frac{y_\ell}{\pi_\ell p(\mathbf{x}_\ell)}
	-
	\sum_{k\in S_r}
	(1-\pi_k)\,
	\frac{1-p(\mathbf{x}_k)}{p(\mathbf{x}_k)^2}\,
	\frac{y_k^2}{\pi_k^2}
	\right\} \notag \\
	&\quad
	+
	\frac{1}{N^2}
	\sum_{k\in S_r}
	w_k^2\,
	\frac{1-p(\mathbf{x}_k)}{p(\mathbf{x}_k)^2}\,
	y_k^2 .
	\label{eq:Vhat_tot_IPW}
\end{align}
The variance estimator
\eqref{eq:Vhat_tot_IPW} is $dq$-unbiased in the sense that
$
\E_d\E_q\!\left\{
\widehat{\V}\!\left(\widehat{\mu}_{\mathrm{ipw}}(p)\right)
\right\}
=
\V_{dq}\!\left\{\widehat{\mu}_{\mathrm{ipw}}(p)\right\}.
$
Moreover, under standard regularity conditions for two-phase sampling, the oracle IPW estimator
$\widehat{\mu}_{\mathrm{ipw}}(p)$ is root-$n$ consistent and asymptotically
normal (e.g., \cite{chen2007asymptotic}) under the $dq$ framework, and the variance estimator
\eqref{eq:Vhat_tot_IPW} is consistent for $\V_{dq}\!\left\{\widehat{\mu}_{\mathrm{ipw}}(p)\right\}$. An approximate $(1-\alpha)$ Wald-type confidence interval for $\mu$ is thus given by
\[
\widehat{\mu}_{\mathrm{ipw}}(p)
\;\pm\;
z_{1-\alpha/2}\,
\left[
\widehat{\V}\!\left\{\widehat{\mu}_{\mathrm{ipw}}(p)\right\}
\right]^{1/2}.
\]

\subsection{THE FEASIBLE IPW ESTIMATOR}
In practice, the response probability function $p(\cdot)$ is unknown
and must be estimated from the sample data.
Let $\widehat p(\cdot)$ denote an estimator of $p(\cdot)$
constructed using the observed sample. 
Here, $p(\cdot)$ plays the role of a nuisance function: it is not of
primary interest but is introduced only to construct the IPW estimator
of the population mean. The resulting feasible IPW estimator of  $\mu$ is
\begin{equation}\label{eq:IPW_feasible_mean}
	\widehat{\mu}_{\mathrm{ipw}}(\widehat p)
	=
	\frac{1}{N}\sum_{k\in S_r}
	\frac{w_k}{\widehat p(\mathbf{x}_k)}\,y_k
	=
	\frac{1}{N}\sum_{k\in S_r}
	w_k^*\,y_k,
\end{equation}
where $w_k^* = w_k / \widehat p(\mathbf{x}_k)$ denotes the
response-adjusted weight associated with unit $k$.
The collection $\{w_k^* : k \in S_r\}$ defines the response-adjusted weighting
system, which is applied to all survey variables of interest
$Y_1,\ldots,Y_p$.\\

\noindent
To understand the effect of estimating the response probabilities, it is
useful to compare the feasible estimator with the oracle IPW estimator.
The feasible estimator can be written as
\begin{align}\label{eq:IPW_decomposition}
	\widehat{\mu}_{\mathrm{ipw}}(\widehat p)
	=
	\widehat{\mu}_{\mathrm{ipw}}(p)
	+
	\frac{1}{N}\sum_{k\in S_r}
	w_k y_k
	\left\{
	\frac{1}{\widehat p(\mathbf{x}_k)}
	-
	\frac{1}{p(\mathbf{x}_k)}
	\right\}.
\end{align}

\noindent
This decomposition shows that the difference between the feasible and oracle
estimators is driven by the estimation error in the response probabilities,
through the remainder term on the right-hand side of
\eqref{eq:IPW_decomposition}\\

\noindent
Consider first the case where the response probabilities are estimated
using a correctly specified parametric model
$p(\mathbf{x};\boldsymbol{\alpha})$.
Under the positivity assumption and mild smoothness conditions, a Taylor
expansion of the function $u\mapsto u^{-1}$ around $u=p(\mathbf{x}_k)$ yields
\[
\frac{1}{\widehat p(\mathbf{x}_k)}
-
\frac{1}{p(\mathbf{x}_k)}
=
-
\frac{\widehat p(\mathbf{x}_k)-p(\mathbf{x}_k)}{p(\mathbf{x}_k)^2}
+
R_k,
\]
where, by Taylor's theorem,
\[
R_k
=
\frac{\{\widehat p(\mathbf{x}_k)-p(\mathbf{x}_k)\}^2}{\xi_k^3},
\]
for some $\xi_k$ lying between $\widehat p(\mathbf{x}_k)$ and
$p(\mathbf{x}_k)$.
Substituting this expansion into \eqref{eq:IPW_decomposition} yields
\begin{align}\label{eq:IPW_Taylor}
	\widehat{\mu}_{\mathrm{ipw}}\!\left(\widehat p\right)
	&=
	\widehat{\mu}_{\mathrm{ipw}}(p)
	-
	\frac{1}{N}\sum_{k\in S_r}
	w_k\,y_k\,
	\frac{\widehat p(\mathbf{x}_k)-p(\mathbf{x}_k)}{p(\mathbf{x}_k)^2}
	+
	\frac{1}{N}\sum_{k\in S_r}
	w_k\,y_k\,R_k.
\end{align}
Decomposition~\eqref{eq:IPW_Taylor} clarifies the role of response probability
estimation in the asymptotic behavior of the IPW estimator. The leading
correction term is linear in the estimation error
$\widehat p(\mathbf{x}_k)-p(\mathbf{x}_k)$, while the remainder term involves
its squared magnitude. \\

\noindent
When a correctly specified parametric model
$p(\mathbf{x};\boldsymbol{\alpha})$ is used and standard regularity conditions
hold, the estimator $\widehat{\boldsymbol{\alpha}}$ is root-$n$ consistent.
As a result, the linear term contributes at the $n^{-1/2}$ scale, whereas the
remainder term is asymptotically negligible. 
As a simple parametric example, the response probabilities may be
modeled using logistic regression. Specifically, one may assume
\[
p(\mathbf{x}_k;\boldsymbol{\alpha})
=
\frac{\exp(\boldsymbol{\alpha}^\top \mathbf{x}_k)}
{1+\exp(\boldsymbol{\alpha}^\top \mathbf{x}_k)},
\]
where the parameter vector $\boldsymbol{\alpha}$ is estimated from the
sample by maximizing a (possibly design-weighted) pseudo-likelihood
based on $\{(r_k,\mathbf{x}_k):k\in S\}$; see, for example, \cite{haziza2017construction}. The fitted response probabilities
$\widehat p(\mathbf{x}_k)=p(\mathbf{x}_k;\widehat{\boldsymbol{\alpha}})$
are then used to construct the feasible IPW estimator. Under standard regularity conditions, the resulting estimator
$\widehat{\mu}_{\mathrm{ipw}}(\widehat p)$ is root-$n$ consistent for
$\mu$ and asymptotically normal under the $dq$ framework. Because the
IPW estimating equation is not Neyman orthogonal with respect to the
response probability function $p(\cdot)$ (see Section S1 of the Supplementary Material), the estimation of
$\boldsymbol{\alpha}$ contributes at first order to the asymptotic
distribution of the estimator. Consequently,
$\widehat{\mu}_{\mathrm{ipw}}(\widehat p)$ is generally not
asymptotically equivalent to the oracle estimator
$\widehat{\mu}_{\mathrm{ipw}}(p)$, as its asymptotic variance contains
an additional component induced by estimating
$\boldsymbol{\alpha}$. This lack of orthogonality also complicates variance estimation. In
classical parametric or low-dimensional settings, this is usually not a
major issue, as standard linearization methods can be derived by
explicitly accounting for the estimation of the response probabilities;
see, for example, \cite{kim2007nonresponse}.\\

\noindent  Parametric models for the response probabilities can be sensitive to
misspecification. If important nonlinear effects or interactions are omitted,
the estimated response propensities may be unreliable and the resulting IPW
estimators inconsistent. Another practical difficulty is that very small
estimated response probabilities can produce extremely large weights and
unstable estimators. For these reasons, national statistical offices often use
more flexible methods that avoid specifying a particular functional form.\\

\noindent A method widely used in survey practice is propensity score
stratification, also known as the score method. Preliminary response
propensities, often obtained from a parametric model such as logistic
regression, are used to form groups of units with similar response
probabilities. Within each group, the nonresponse adjustment is based on the
observed response rate, which helps reduce nonresponse bias while limiting the
variability of the weights. This approach has a long history in survey
methodology; see, for example, \cite{little1986survey, el1997diagnostics} and
\cite{haziza2007construction}. Closely related stratification methods based on
propensity scores have also been studied in causal inference; see, for example,
\cite{lunceford2004stratification}. More flexible nonparametric approaches
have also been proposed: \cite{silva2006kernel} consider kernel smoothing
methods, while \cite{da2009nonparametric} extend this framework using local
polynomial regression. These methods avoid specifying a parametric link
function but suffer from the curse of dimensionality when the number of
auxiliary variables is moderate to large. More recently,
\cite{opsomer2025schaid} develop a design-based recursive partitioning method
(sCHAID) for constructing nonresponse adjustment cells. Regression trees have
also been studied for modeling response propensities; e.g.,
\cite{phipps2012analyzing}. Other ML methods, such as random forests or
boosting procedures, have been explored in recent work. Most of these
contributions rely on simulation studies or empirical comparisons to assess
the impact of ML-based propensity estimation on IPW estimators; see, for
example, \cite{lohr2015using}, \cite{kern2019tree}, and \cite{larbi2025use}.
While these studies provide useful empirical insights, a comprehensive
theoretical understanding of ML-based propensity estimation in survey settings
remains limited.\\

\noindent Many ML estimators of $p(\cdot)$ converge at rates slower than $n^{-1/2}$.
In that case, the quadratic remainder in \eqref{eq:IPW_Taylor} may no
longer be negligible. Moreover, because many ML estimators are highly
data-adaptive and non-smooth as functions of the sample, standard
linearization arguments become difficult to apply. When this occurs,
root-$n$ consistency may fail, the usual central limit theorem may break
down, and standard inference procedures can become unreliable.\\

\noindent These challenges naturally raise the question of whether the DML
framework could help. As discussed earlier for imputation
(see Section~\ref{Sec:IMP}), DML can allow the use of flexible ML methods
while preserving root-$n$ inference through Neyman-orthogonal estimating
equations. However, extending this idea to survey nonresponse is not
straightforward. DML is typically built around outcome-specific estimating
equations, such as the AIPW estimator, which must be constructed separately
for each outcome variable. In contrast, standard survey practice aims to
construct a single set of nonresponse-adjusted weights that can be applied to
all survey variables. Because AIPW estimators depend explicitly on the outcome
variable $Y$, they do not naturally fit into this framework. Thus, DML methods
are outcome-specific, whereas survey weighting is designed to be
outcome-agnostic. Reconciling these two perspectives requires further
methodological development.\\

\noindent
In contrast with the parametric setting, variance estimation becomes
substantially more challenging when the response model is estimated
using flexible ML procedures. In general, standard linearization
techniques are no longer readily available, as the estimators of
$p(\cdot)$ may be highly adaptive and lack the smoothness required for
classical Taylor expansions. As a result, resampling methods such as the
bootstrap may fail to capture the true variability of IPW estimators,
especially when the estimated weights are unstable. This can occur when
some estimated response probabilities are very small or when the ML
estimator of $p(\cdot)$ is highly non-smooth. That said, bootstrap failure is not inherent to ML itself. When the
response probability estimator is reasonably stable—so that small changes
in the data lead to small changes in the fitted probabilities—the
bootstrap may still provide useful finite-sample approximations.
However, as the learning algorithm becomes more adaptive, and in the
absence of Neyman orthogonality, the reliability of bootstrap-based
inference becomes increasingly uncertain. In Section~S3 of the
Supplementary Material, we illustrate the use of a pseudo-population
bootstrap procedure (e.g., \cite{mashreghi2016survey}) for IPW estimation and examine its finite-sample
performance for several ML methods used to estimate the response
probabilities.\\

\noindent Finally, the choice of model architecture for estimating response probabilities
raises a fundamental issue. In ML, model selection is typically driven by predictive performance, for
example through cross-validation aimed at minimizing a prediction loss for
$\widehat p(\mathbf{x})$. However,
minimizing the prediction error of the response model does not necessarily
minimize the mean squared error of the resulting IPW estimator. A response model
that achieves good predictive accuracy may still yield highly variable
response-adjusted weights $w_k^*$ and an unstable estimator of $\mu$. In particular, including auxiliary variables that are highly predictive of response but unrelated to the survey variable does not reduce nonresponse bias, yet may substantially inflate variance. Such variables can induce small	estimated response probabilities, leading to highly dispersed weights and unstable estimators, for reasons analogous to Basu’s paradox \citep{basu1971essay}. As a result,
variables that are predictive of response but weakly or not related to the
survey outcome may increase variance without any corresponding bias reduction
and are often best excluded from the response model \cite{little2005does, beaumont2005use, park2019note}. From an inferential perspective, the relevant objective is therefore not prediction accuracy per se, but rather control of the mean squared error of the
IPW estimator itself. Determining how to select an appropriate model
architecture and an effective criterion for doing so remains an important
topic for further research.\\

\noindent Figure~\ref{Fig2} reports side-by-side boxplots of the Monte Carlo percent relative error of the IPW estimator under different machine learning architectures. The full simulation study is presented in Section S2 of the Supplementary Material. The main message from this figure is that the performance of the IPW estimator can vary substantially with the choice of architecture and tuning parameters. In some cases, the Monte Carlo percent relative errors are nearly centered around zero, whereas in others, sizeable biases are observed. This should not be interpreted as evidence that machine learning methods are not useful for inverse probability weighting. Rather, it highlights the need for care when selecting the learning algorithm, its architecture, and its hyperparameters, since these choices may have a non-negligible impact on the resulting estimator.

\begin{center}
	\begin{figure}[H]
		\centering
		\includegraphics[width=0.95\linewidth]{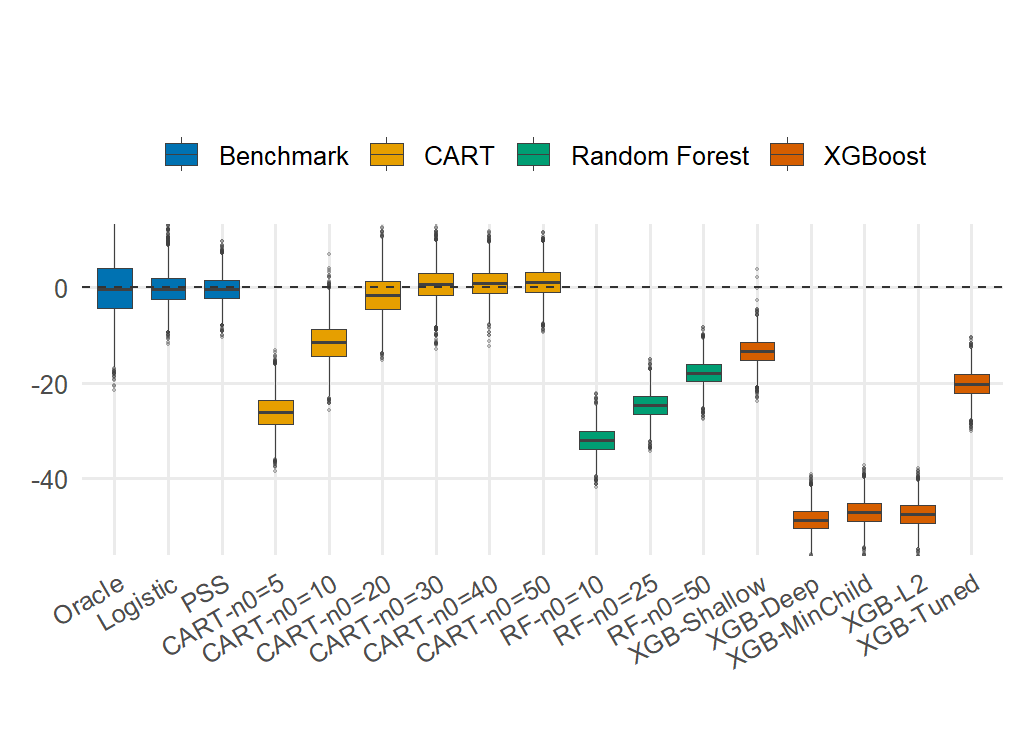}
		\caption{Distribution of the percent relative error of $\widehat{\mu}_{\mathrm{ipw}}(\widehat p)$ based on several ML methods.}
		\label{Fig2}
	\end{figure}
\end{center}

\section{FINAL REMARKS}
In this final section, we briefly discuss two areas in which the use of ML methods raises important methodological questions: small area estimation and the integration of probability and non-probability samples.

\subsection{SMALL AREA ESTIMATION}

Another important problem in official statistics is that of small area estimation (SAE). 
The goal is to produce reliable estimates for small subpopulations or geographic domains 
(e.g., regions, municipalities, demographic groups), for which the available sample sizes 
are often very limited or even zero (e.g., \cite{rao2015small}). As a result, direct design-based estimators tend to be 
highly variable or unusable in practice.\\

\noindent To address this issue, SAE methods rely on statistical models that allow one to 
``borrow strength'' across domains or from auxiliary information. In particular, linear 
mixed models are widely used to combine information from multiple areas and improve 
estimation accuracy in the presence of small samples. Two main modeling frameworks are typically distinguished: area-level models and  unit-level models. Area-level models are based on aggregated data (e.g., direct estimators 
and their sampling variances), whereas unit-level models rely on individual-level data and 
explicit modeling of the outcome variable. While both approaches share the same objective, 
they differ substantially in their statistical structure and in the type of information 
available.\\

\noindent The connection with machine learning arises naturally, as SAE is fundamentally a 
prediction problem. A growing literature has explored the use of ML methods in this context,
including tree-based, random-forest, and mixed-effects approaches; see, for
example, \cite{krennmair2022flexible, michal2024model,
	frink2024small, parker2026, krennmair2026random}. Despite these advances, a general framework for using ML methods in SAE while
retaining valid uncertainty quantification remains less developed. We believe that developing such a framework is an important and promising 
direction for future research, and this is currently the subject of ongoing work.

\subsection{DATA INTEGRATION OF PROBABILITY AND NONPROBABILITY SAMPLES}
Data integration has become an important topic in survey statistics due to the increasing availability of alternative data sources and the growing challenges faced by traditional surveys. A common framework for data integration considers the combination of a probability sample (denoted by $S_P$) and a non-probability sample (denoted by $S_{NP}$). For units in the probability sample $S_P$, auxiliary variables $\mathbf{x}_k$ are observed together with the sampling weights $w_k=1/\pi_k$, but the survey variable $y_k$ may not be available. In contrast, for units in the non-probability sample $S_{NP}$, both the survey variable $y_k$ and the auxiliary variables $\mathbf{x}_k$ are observed, but the inclusion mechanism in the sample $S_{NP}$ is unknown. The goal is to combine the information from these two sources in order to estimate population quantities such as the finite population mean of $Y$.\\

\noindent Several approaches have been proposed in the literature to address this problem. One class of methods relies on propensity score weighting, where the probability of inclusion in the non-probability sample is modeled as a function of the auxiliary variables and used to construct weights that correct for selection bias. A second class of methods is based on mass imputation, where the relationship between $Y$ and $\mathbf{x}$ is estimated using the non-probability sample and then used to impute values of $Y$ for units in the probability sample. More recently, doubly robust estimators have been proposed, combining both approaches and remaining consistent if either the propensity score model or the outcome model is correctly specified. Important contributions in this direction include the doubly robust framework proposed by \cite{chen2020doubly}, which develops estimators for combining probability and non-probability samples under parametric models for both the propensity score and the outcome regression functions.\\

\noindent In recent years, ML methods have been increasingly used to estimate the nuisance functions involved in these procedures, such as the selection probability or the outcome regression function. These flexible data-adaptive methods can help mitigate the risk of model misspecification when simple parametric models are inadequate. A number of recent studies have explored the use of ML techniques in the context of data integration, including the use of classification and regression trees, random forests, or boosting methods to estimate propensity scores or outcome models (e.g., \cite{ferri2020propensity, ferri2022weight, ferri2024estimating, beaumont2024handling}). These approaches aim to improve robustness to model misspecification and to better capture complex relationships between the study variable and auxiliary variables. However, the use of ML in this context raises important questions regarding statistical inference. In particular, standard estimators may inherit the slow convergence rates of ML estimators, which can lead to non-negligible bias and invalid confidence intervals.\\	

\noindent As noted by \cite{seaman2025debiased}, the existing literature on
data integration has largely focused on point estimation and has paid
comparatively little attention to the validity of statistical inference when
ML estimators are used for the nuisance functions. To address this issue,
\cite{seaman2025debiased} proposed a framework based on DML, which combines
doubly robust estimators with cross-fitting to obtain asymptotically normal
estimators even when flexible ML methods are used. To the best of our
knowledge, this work is among the first to provide a theoretical justification
specifically for DML methods with cross-fitting and data-adaptive nuisance
estimation in the context of data integration combining probability and
non-probability samples.

\bibliographystyle{apalike}
\bibliography{biblio}
\addcontentsline{toc}{section}{References}
\newpage

\appendix

\section{A digression on modern semiparametric theory} \label{appendix1}

In this section of the supplement, we introduce several concepts that will be useful for understanding the notion of orthogonality, which appears repeatedly throughout the paper and plays a central role in the DML framework.

\noindent Consider a classical semiparametric setting where we observe  $\mathbf{z}_1, \ldots, \mathbf{z}_n  \overset{\text{i.i.d.}}{\sim} \P_{\theta_\star, \eta_\star}$ parametrized by a parameter of interest $\theta_\star\in \Theta \subseteq\mathbb{R}^q$  and a nuisance parameter $\eta_\star\in \mathcal{H}$, with a nuisance space $\mathcal{H}$ infinite dimensional. Typically, $\eta_\star$ represents a function and thus $\mathcal{H}$ is a function space; for instance,	if $\mathbf{z}=(\mathbf{x}^\top,y)^\top$, then
$\eta_\star:\mathbf{x}\mapsto
\E[y\mid\mathbf{X}=\mathbf{x}]$.

\noindent The goal would be to estimate the finite-dimensional parameter of interest $\theta_\star$. Specifically, we would like to construct an estimator $\widehat{\theta}_n$ such that
$$\sqrt{n}(\widehat{\theta}_n - \theta_\star)\xrightarrow[n \to \infty]{\mathcal{L}}\mathcal{N}(0, V)$$ 
for some asymptotic variance $V$.   However, estimation of the nuisance parameter $\eta_\star$ is often necessary and thus makes the problem more challenging. Indeed, given that $\mathcal{H}$ is infinite-dimensional, we typically have $\rVert \widehat{\eta}_n - \eta_\star\rVert = o_\P(1)$ for the $L^2$ norm under fairly weak conditions but not $\rVert \widehat{\eta}_n - \eta_\star\rVert = O_\P(n^{-1/2})$; that is, slower rates of convergence. 

\noindent Commonly, the parameter of interest $\theta_\star$ can be identified as the solution of an estimating equation $$U(\theta_\star, \eta_\star) := \E \left[ \psi \left(\mathbf{z}; \theta_\star, \eta_\star\right)\right]=0,$$ for some function $\psi$. A classical approach would be a two-step process: (1) estimate $\eta_\star$ on the observed data with an estimator $\widehat{\eta}_n$; (2) plug in $\widehat{\eta}_n$ to define the estimator $\widehat{\theta}_n$ of $\theta_\star$ by solving for $\theta$ the sample moment condition $$ \dfrac{1}{n}\sum_{k=1}^n \psi \left(\mathbf{z}_k, \theta, \widehat{\eta}_n\right) = 0.$$ 
Then, assuming appropriate smoothness conditions on $\psi$ in both $\theta$ and $\eta$ and consistency of $\widehat{\eta}_n \to \eta_\star$, a pedagogical decomposition gives
{\small
	\begin{align*}
		\widehat{\theta}_n - \theta_\star
		&= -\boldsymbol{G}_\theta^{-1} \bigg\{
		\underbrace{\dfrac{1}{n}\sum_{k=1}^n \psi(\mathbf{z}_k, \theta_\star, \eta_\star)}_\text{(a)}
		+ \underbrace{D_\eta U(\theta_\star, \eta_\star)\big[\widehat{\eta}_n - \eta_\star\big]}_\text{(b)}
		\\&+ \underbrace{\dfrac{1}{n}\sum_{k=1}^n \Big(
			\psi(\mathbf{z}_k, \theta_\star, \widehat{\eta}_n)
			- \psi(\mathbf{z}_k, \theta_\star, \eta_\star)
			\Big)
			- \mathds{E}\Big[
			\psi(\mathbf{z}, \theta_\star, \widehat{\eta}_n)
			- \psi(\mathbf{z}, \theta_\star, \eta_\star)
			\Big]}_\text{(c)} \\&+ \underbrace{\mathcal{O}_\P \left(\rVert \widehat{\eta}_n - \eta_\star \rVert^2\right)}_\text{(d)}
		\bigg\} + o_\P(n^{-1/2}),
\end{align*}}
where $$\boldsymbol{G}_\theta := \left.
\frac{\partial}{\partial \theta^\top}
\mathbb{E}\!\left[\psi(\mathbf{z};\theta,\eta_\star)\right]
\right|_{\theta=\theta_\star}$$ denotes the Jacobian in $\theta$ and $$D_\eta U(\theta_\star,\eta_\star)[h]
= \lim_{t \to 0} t^{-1}(
\mathbb{E}[\psi(\mathbf{z};\theta_\star,\eta_\star + t h)]
- \mathbb{E}[\psi(\mathbf{z};\theta_\star,\eta_\star)])
$$ 
represents the Gateaux derivative in direction $h$. The first term (a) is well-behaved and, under appropriate moment conditions, will be asymptotically normal when rescaled by $\sqrt{n}$. The second and the third terms (b) and (c) may become problematic. The Double Machine Learning (DML) methodology suggested by Chernozhukov et~al. (2018) is tailored to treat these terms. Specifically, the second term may, in general, not be well-behaved. One of the ingredients of the DML methodology is the use of \emph{Neyman orthogonal} estimating equations: an estimating equation is said to be Neyman orthogonal if $$D_\eta U(\theta_\star, \eta_\star)[\eta - \eta_\star]=0$$ for every admissible direction $h$. If the estimating equation is Neyman orthogonal, then (b)$=0$. The third term can be shown to be negligible by using cross-fitting. The idea of cross-fitting is to split the data (in half, say; assume $n$ is even), use one part of the sample (the first, say) to fit $\widehat{\eta}_{n/2}^{(1)}$, then use the other (the second) to solve $$\sum_{k =n/2+1}^n \psi (\mathbf{z}_k, \theta, \widehat{\eta}_{n/2}^{(1)}) = 0$$ for $\theta$ to get $\widehat{\theta}_{n/2}^{(1)}$. The roles are then reversed to produce   $\widehat{\theta}_{n/2}^{(2)}$. Finally, the two estimators are averaged to get 
$$\widehat{\theta}_n = (\widehat{\theta}_{n/2}^{(1)}+\widehat{\theta}_{n/2}^{(2)})/2.$$ 
The key idea here is that, conditional on the training fold, we may see (c) as an average of conditionally i.i.d. mean-zero random variables (within evaluation folds) with conditional variance shrinking as $\rVert\widehat{\eta}_n - \eta_\star \rVert^2 \to 0$. Therefore, under suitable smoothness conditions, the overall order of that term is $O_\P (n^{-1/2} \rVert \widehat{\eta}_n - \eta_\star \rVert)$ which is asymptotically negligible when rescaled by $\sqrt{n}$. Finally, term (d) is a term which can be made asymptotically negligible if the nuisance parameters are estimated fast enough, specifically, if $\rVert\widehat{\eta}_n - \eta_\star \rVert = o_\P(n^{-1/4})$. This is still a critical assumption which requires good convergence properties of $\widehat{\eta}_n $, but much weaker than assuming a parametric rate. Overall, this deliberately simplified expansion shows that, under suitable smoothness and rate conditions, the resulting DML estimator may be asymptotically equivalent to the oracle estimator that treats $\eta_\star$ as known. \\

\noindent 

The above discussion aims to provide an informal overview of the DML methodology in a classical i.i.d. setting. While the objectives in survey sampling are closely related, they are not identical, and adapting DML ideas to survey sampling problems requires particular care. For example, in many sampling designs, the sampling indicators $(I_k)_{k\in U}$ introduced earlier are neither independent nor identically distributed, so key ingredients of the standard DML framework such as cross-fitting and conditional independence arguments do not directly transpose. As a result, the use of DML in survey sampling often requires  problem-specific modifications. The aim of this paper is precisely to review, in a pedagogical manner, how such adaptations can be carried out in the main survey sampling settings where modeling arises.

\subsection{The model-assisted estimator}

Let $\mathbf{z}_k := (I_k, y_k, \bx_k)$ for $k \in U$. The finite population estimating equation leading to the model-assisted estimator is given by $$\psi_{\rm MA}(\mathbf{z}_k, \theta, f) := \dfrac{I_k}{\pi_k}\left(y_k - f(\bx_k)\right) + f(\bx_k)-\theta.$$
Indeed, for any $f: \mathbb{R}^p\to\mathbb{R}$, $$ U_{\rm MA}( \mu, f)= \mathds{E}_d \left[ \dfrac{1}{N}\sum_{k\in U}\psi_{\rm MA}(\mathbf{z}_k, \mu, f)\right] = 0, \qquad  \dfrac{1}{N}\sum_{k\in U}\psi_{\rm MA}(\mathbf{z}_k,  \widehat{\mu}_{\rm MA}(f), f) =0.$$
Moreover, $$D_f U_{\rm MA}(\mu, f)[h] = \dfrac{1}{N}\sum_{k\in U} \E_d \left[\left(1 - \dfrac{I_k}{\pi_k}\right) h(\bx_k)\right] = 0, \qquad \forall h:\R^p \to \R.$$
This implies Neyman orthogonality. 
\subsection{The customary imputed estimator}
Let $\mathbf{z}_k := (I_k, r_k, y_k, \bx_k)$ for $k \in U$. The finite population estimating equation leading to the imputed estimator is given by $$\psi_{\rm imp}(\mathbf{z}_k, \theta, f) := \dfrac{I_k}{\pi_k}\left(r_ky_k+(1-r_k)f(\bx_k)\right) -\theta.$$
Indeed, for $m: \bx \mapsto \E[y_k \rvert \bx_k=\bx]$, $$ U_{\rm imp}( \mu, m)= \mathds{E}_{mdq} \left[ \dfrac{1}{N}\sum_{k\in U}\psi_{\rm imp}(\mathbf{z}_k, \mu, m)\right] = 0, \qquad  \dfrac{1}{N}\sum_{k\in U}\psi_{\rm imp}(\mathbf{z}_k,  \widehat{\mu}_{I}(f), f) =0.$$
However, for a perturbation $h$, $$D_m U_{\rm imp}(\mu, m)[h] = \dfrac{1}{N}\sum_{k\in U}(1-p(\bx_k))h(\bx_k)$$ which is typically not equal to zero. 
\subsection{The AIPW estimator}
Let $\mathbf{z}_k := (I_k, r_k, y_k, \bx_k)$ for $k \in U$. The finite population estimating equation leading to the AIPW estimator is given by $$\psi_{\rm aipw}(\mathbf{z}_k, \theta, f,g) := \dfrac{I_k}{\pi_k}\left(f(\bx_k) +\dfrac{r_k (y_k-f(\bx_k))}{g(\bx_k)} \right) -\theta.$$
Indeed, for $m: \bx \mapsto \E[y_k \rvert \bx_k=\bx]$ and $p: \bx \mapsto \P(r_k=1\rvert \bx_k=\bx)$, $$ U_{\rm aipw}( \mu, m, p)= \mathds{E}_{mdq} \left[ \dfrac{1}{N}\sum_{k\in U}\psi_{\rm aipw}(\mathbf{z}_k, \mu, m, p)\right] = 0, \qquad  \dfrac{1}{N}\sum_{k\in U}\psi_{\rm aipw}(\mathbf{z}_k,  \widehat{\mu}_{\rm aipw}(f,g), f,g) =0.$$
For perturbations $h_1$ and $h_2$,
\begin{align*}
	D_{(f,g)}U_{\mathrm{aipw}}(\mu,m,p)[h_1,h_2]
	&=
	\frac{1}{N}\sum_{k\in U}
	\E_m\E_d\E_q\left[
	\frac{I_k}{\pi_k}
	\left\{
	\left(1-\frac{r_k}{p(\bx_k)}\right)h_1(\bx_k)
	-
	\frac{r_k\{y_k-m(\bx_k)\}}{p(\bx_k)^2}h_2(\bx_k)
	\right\}
	\right] \\
	&=0.
\end{align*} 
This shows Neyman orthogonality of the AIPW estimating equation.
\subsection{The IPW estimator}

Let $\mathbf{z}_k := (I_k, r_k, y_k, \bx_k)$ for $k \in U$. The finite population estimating equation leading to the IPW estimator is given by $$\psi_{\rm ipw}(\mathbf{z}_k, \theta, f) := \dfrac{I_kr_ky_k}{\pi_kf(\bx_k)} -\theta.$$
Indeed, for $p: \bx \mapsto \P(r_k=1\rvert \bx_k=\bx)$, $$ U_{\rm ipw}( \mu, p)= \mathds{E}_{dq} \left[ \dfrac{1}{N}\sum_{k\in U}\psi_{\rm ipw}(\mathbf{z}_k, \mu, p)\right] = 0, \qquad  \dfrac{1}{N}\sum_{k\in U}\psi_{\rm ipw}(\mathbf{z}_k,  \widehat{\mu}_{\rm ipw}(f), f) =0.$$
However, for a perturbation $h$, $$D_d U_{\rm ipw}(\mu, p)[h] = -\dfrac{1}{N}\sum_{k\in U}\dfrac{y_kh(\bx_k)}{p(\bx_k)}$$ which is typically not equal to zero.

\section{Simulation study: Choice of the architecture}\label{sec:simulation}

We conducted a simulation study to assess the impact of the architecture of
machine learning procedures on the behavior of the inverse probability
weighted (IPW) estimator in the presence of unit nonresponse. The setting was
deliberately chosen to be favorable: the nonresponse mechanism is a
main-effects logistic model, correctly specified in the available covariates,
so that a standard parametric approach is expected to perform well. The goal
is to document the extent to which flexible learning procedures, fitted with
different but seemingly reasonable hyperparameter configurations, can
nevertheless lead to markedly different estimators of a finite population
mean.


We first generated a finite population of size $N=20{,}000$. For each unit
$k=1,\dots,N$, a vector of $p=10$ covariates
$\mathbf{x}_k=(x_{k1},\dots,x_{k10})^\top$ was generated, whose components
were drawn independently from a Gamma distribution with shape parameter equal
to $1$ and scale parameter equal to $10$, so that
$E(x_{kj})=10$ and $\mathbb{V}(x_{kj})=100$. The survey variable was
generated according to the linear model
\begin{equation}\label{eq:sim-outcome}
	y_k \;=\; \mathbf{x}_k^\top\boldsymbol{\beta} + \varepsilon_k,
	\qquad \varepsilon_k \overset{iid}{\sim} \mathcal{N}(0,\sigma^2),
\end{equation}
with $\boldsymbol{\beta}=(2,\dots,2)^\top$. The variance $\sigma^2$ was set
so that the
population coefficient of determination was equal to $R^2=0.5$. Finally, the
$y$-values were shifted by a constant so that the finite population mean,
$
\mu \;=\; {N}^{-1}\sum_{k\in U} y_k,
$
was exactly equal to $200$; this recentering affects neither the relationship
between the survey variable and the covariates nor the value of $R^2$. The
population, generated once, was kept fixed throughout the simulation: the randomness in the experiment stems from
repeated sampling and repeated nonresponse only.

From the population, we selected $M=10{,}000$ samples, each of size $n=500$,
according to simple random sampling without replacement, leading to a common
design weight $w_k = N/n = 40$ for all $k$.


In each sample, the response indicators $r_k$ were generated independently
from a Bernoulli distribution with probability
\begin{equation}\label{eq:sim-response}
	p(\mathbf{x}_k) \;=\; \Pr\left(r_k = 1 \mid \mathbf{x}_k\right)
	\;=\; \frac{\exp\!\left(\gamma_0 +
		\mathbf{x}_k^\top\boldsymbol{\gamma}\right)}
	{1+\exp\!\left(\gamma_0 + \mathbf{x}_k^\top\boldsymbol{\gamma}\right)},
\end{equation}
with
$\boldsymbol{\gamma} = (0.05,\,-0.04,\,0.03,\,-0.02,\,0.01,\,0.04,\,-0.03,\,
0.02,\,-0.01,\,0.02)^\top.$ The intercept $\gamma_0$ was determined
numerically so that the average response probability in the population was
equal to $50\%$. The mechanism \eqref{eq:sim-response} is a main-effects
logistic model in the covariates $\mathbf{x}_k$; a main-effects logistic
regression fitted to the sample data is therefore correctly specified.


In each sample, we computed the IPW estimator of $\mu$,
\begin{equation}\label{eq:sim-ipw}
	\widehat{\mu}_{\rm ipw}(\widehat{p}) \;=\;
	\frac{1}{N}\sum_{k\in S_r}
	\frac{w_k\, y_k}{\widehat{p}(\mathbf{x}_k)},
\end{equation}
where $S_r$ denotes the set of respondents and
$\widehat{p}(\mathbf{x}_k)$ denotes the estimated response probability
attached to unit $k$. As a benchmark, we also computed the oracle estimator
$\widehat{\mu}_{\rm ipw}(p)$, obtained by replacing
$\widehat{p}(\mathbf{x}_k)$ in \eqref{eq:sim-ipw} with the true response
probability $p(\mathbf{x}_k)$; while infeasible in practice, it provides a
useful reference point. To prevent unduly large weights, the estimated
probabilities were truncated to the interval $[0.001,\,1]$ before computing
\eqref{eq:sim-ipw}.

All the estimation procedures described below were fitted on the full sample
data $\{(r_k,\mathbf{x}_k);\ k\in S\}$ and the fitted values
$\widehat{p}(\mathbf{x}_k)$ were obtained for the same units, mirroring
common practice, whereby the respondents of a given sample are reweighted
using a model fitted on that same sample.


Because $p(\mathbf{x}) = E(r \mid \mathbf{x})$ is a conditional expectation,
we treated the estimation of the response probabilities as a regression
problem: all the tree-based procedures below were fitted with the
least-squares criterion, treating the response indicator as a continuous
variable. In addition to the oracle, we considered the following procedures:

\begin{enumerate}
	\item \textbf{Logistic regression (LR).} A main-effects logistic
	regression of $r_k$ on $\mathbf{x}_k$, which coincides with the true
	nonresponse mechanism \eqref{eq:sim-response}.
	
	\item \textbf{Propensity score stratification (PSS).} The units were
	partitioned into $G=10$ classes based on the deciles of the scores
	obtained from the logistic regression in (i). Within each class, the
	estimated response probability was set equal to the observed response
	rate.
	
	\item \textbf{Regression trees (CART).} Fully grown regression trees
	fitted with the {\tt rpart} package with no cost-complexity penalty
	(${\tt cp}=0$), so that the complexity of the partition is governed
	solely by the minimum number of observations in the terminal nodes,
	denoted by $n_0$. We considered six values,
	$n_0 \in \{5, 10, 20, 30, 40, 50\}$, ranging from very flexible to
	heavily smoothed partitions. The estimated probability of a unit is the
	average of the response indicators in its terminal node. 
	
	\item \textbf{Random forests (RF).} Regression forests of $500$ trees
	fitted with the {\tt randomForest} package, with the number of
	candidate covariates at each split left at its default value
	($\lfloor p/3\rfloor$). As for CART, the complexity was governed by the
	minimum node size, with $n_0 \in \{10, 25, 50\}$. The estimated
	probability of a unit is the average of the predictions of the $500$
	trees.
	
	\item \textbf{Gradient tree boosting (XGB).} Boosting with squared error
	loss, fitted with the {\tt xgboost} package with a subsampling rate of
	$0.8$. We considered the five hyperparameter configurations displayed
	in Table~\ref{tab:sim-xgb}: two contrasted baseline configurations
	(shallow and deep), and three configurations designed as attempts to
	improve on the deep configuration by activating one regularization
	mechanism at a time. We note that, with the squared error loss, the
	{\tt min\_child\_weight} parameter effectively corresponds to a minimum number of observations per terminal node, and hence plays the same role
	as $n_0$ in CART and RF, whereas $\lambda$ is an $L_2$ penalty shrinking
	the predictions of the terminal nodes toward zero. Since squared error
	boosting does not constrain the predictions to the unit interval, the
	fitted values were truncated to $[0,1]$.
\end{enumerate}

\begin{table}[t]
	\centering
	\caption{Hyperparameter configurations for gradient tree boosting.
		All configurations use squared error loss and a subsampling rate
		of $0.8$. Parameters not shown are left at their default values.}
	\label{tab:sim-xgb}
	\begin{tabular}{@{}lrrrrr@{}}
		\toprule
		\textbf{Configuration} & \textbf{Depth} & \textbf{Rounds} &
		$\boldsymbol{\eta}$ & {\tt min\_child\_weight} &
		$\boldsymbol{\lambda}$ \\
		\midrule
		XGB--Shallow   & 2 & 50   & 0.10 & 1  & 1  \\
		XGB--Deep      & 8 & 300  & 0.10 & 1  & 1  \\
		XGB--MinChild  & 8 & 300  & 0.10 & 20 & 1  \\
		XGB--L2        & 8 & 300  & 0.10 & 1  & 50 \\
		XGB--Tuned     & 3 & 100  & 0.05 & 10 & 1  \\
		\bottomrule
	\end{tabular}
\end{table}

In total, $17$ estimators of $\mu$ were computed in each sample: the oracle,
LR, PSS, six versions of CART, three versions of RF, and five versions of
XGB.

\noindent For a given simulation setting, let $\widehat{\mu}^{(m)}$ denote the IPW estimate obtained in Monte Carlo replication $m$, $m=1,\ldots, M$, and let $\mu$ denote the true finite-population mean. The Monte Carlo percent relative error in replication $b$ is defined as
\[
\operatorname{RE}^{(m)}(\widehat{\mu})
=
100\,
\frac{\widehat{\mu}^{(m)}-\mu}{\mu}.
\]
Figure~\ref{Fig2} displays the distribution of
$\operatorname{RE}^{(m)}(\widehat{\mu})$, $m=1,\ldots, M$, across the Monte Carlo replications. Negative values indicate underestimation of the finite-population mean in a given replication, whereas positive values indicate overestimation.\\

\noindent Figure~\ref{Fig2} shows the distribution of the Monte Carlo percent relative errors obtained under the different propensity-score estimation procedures. The Oracle, Logistic, and PSS benchmarks are all approximately centered around zero. In contrast, the results obtained with the machine learning methods depend considerably on the architecture and tuning parameters. For CART, the smallest node-size values lead to substantial negative relative errors: the median is approximately $-25\%$ when $n_0=5$ and approximately $-12\%$ when $n_0=10$. The magnitude of the relative error decreases markedly as the node-size parameter increases, and the results are close to zero for $n_0$ between $20$ and $50$.\\

\noindent A similar, although less favorable, pattern is observed for random forests. All three random forest specifications produce negative relative errors, but their magnitude decreases as the node-size parameter increases, from approximately $-32\%$ for $n_0=10$ to approximately $-18\%$ for $n_0=50$. The results for XGBoost are also highly sensitive to the specification. The shallow architecture yields a relative error of roughly $-14\%$, whereas the deeper and more complex specifications produce errors close to $-50\%$. The tuned version improves substantially upon these specifications, although its relative error remains close to $-20\%$.\\

\noindent Overall, the differences between the centers of the boxplots are much larger than their within-architecture dispersion. This suggests that the observed differences are systematic and cannot be attributed solely to a few unusual Monte Carlo replications. In particular, two specifications belonging to the same class of learning algorithms may lead to very different IPW estimators. These results therefore illustrate that it is not sufficient to select a broad learning method such as CART, random forests, or XGBoost; the particular architecture and its tuning parameters can be equally important. They also suggest that an architecture selected solely on the basis of its predictive performance need not be the one that performs best for the resulting IPW estimator.

\begin{center}
	\begin{figure}[H]
		\centering
		\includegraphics[width=0.95\linewidth]{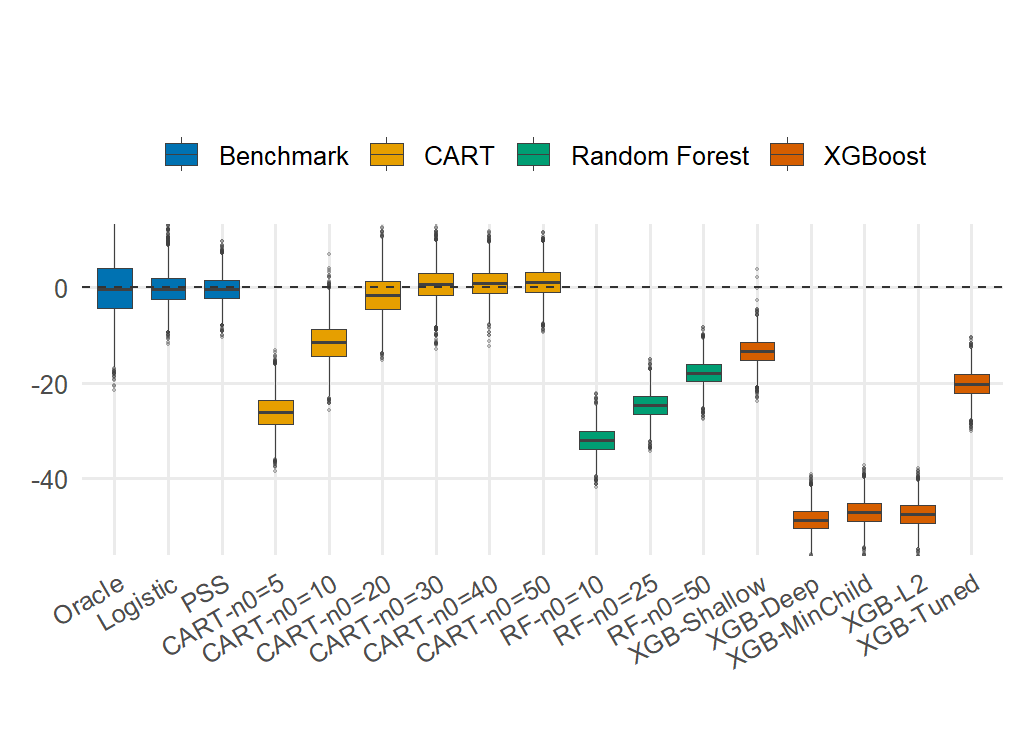}
		\caption{Distribution of the percent relative error of $\widehat{\mu}_{\mathrm{ipw}}(\widehat p)$ based on several ML methods.}
		\label{Fig2}
	\end{figure}
\end{center}

\section{Simulation study: bootstrap variance estimation}

In a separate Monte Carlo simulation study, we examined bootstrap variance
estimation for the Hájek-type IPW estimator. For each scenario, a finite population of size $N$ was generated once and then kept fixed throughout the simulation. Conditional on this finite population, we repeatedly selected simple random samples without replacement and generated nonresponse within each selected sample. The Monte Carlo variation therefore reflects both the sampling design and the nonresponse mechanism, while the underlying finite population remains fixed.

\noindent For each scenario, we considered a finite population of size
\[
N\in\{4000,10000,20000\}.
\]
For each unit $k\in U$, four auxiliary variables were generated independently as
\[
x_{k1}\sim \mathcal{N}(0,1), \qquad
x_{k2}\sim \mathcal{N}(0,1), \qquad
x_{k3}\sim \mathcal{N}(0,1), \qquad
x_{k4}\sim \mathcal{U}(0,1).
\]
The survey variable was generated according to
\[
y_k
=
10 + 2x_{k1} - 1.5x_{k2} + x_{k3} + 3x_{k4} + \varepsilon_k,
\qquad
\varepsilon_k \stackrel{\mathrm{ind}}{\sim} \mathcal{N}(0,2^2),
\qquad k\in U.
\]
For each fixed population, we drew $M=10{,}000$ independent simple random samples without replacement, with
\[
(N,n)\in\{(4000,200),(10000,500),(20000,1000)\}.
\]
Thus, the sampling fraction was fixed at $n/N=0.05$ in all scenarios.

\noindent Within each selected sample, nonresponse was generated independently across sampled units according to a missing-at-random mechanism depending only on the auxiliary variables. More precisely,
\[
r_k\mid \mathbf{x}_k \sim \operatorname{Bernoulli}(p_k),
\qquad
p_k=p(\mathbf{x}_k)=a+(1-a)\operatorname{expit}(\xi_k),
\qquad a=0.1,
\]
where
\begin{align*}
	\xi_k
	&=
	-0.3
	+1.8\,\mathds{1}(x_{k1}>0,\;x_{k2}>0,\;x_{k3}>0)
	-1.5\,\mathds{1}(x_{k1}<-0.5,\;x_{k4}>0.7) \\
	&\quad
	+1.2\,\mathds{1}(x_{k2}>0.5)\,I(x_{k3}<-0.5)
	-0.8\,\mathds{1}(0.3<x_{k4}<0.7)\,I(x_{k1}>0) \\
	&\quad
	+0.6\,\mathds{1}(x_{k1}>1.0)\,I(x_{k2}<-1.0)
	-0.5\,\mathds{1}\!\left\{x_{k3}>\operatorname{med}_{\ell\in U}(x_{\ell3})\right\}\,I(x_{k4}<0.5) \\
	&\quad
	+0.4\,\mathds{1}(|x_{k1}|<0.5)\,I(x_{k2}>0)
	-0.7\,\mathds{1}(x_{k1}>0,\;x_{k2}<0,\;x_{k3}<0).
\end{align*}
Here, $\mathds{1}\{\cdot\}$ denotes the indicator function and $\operatorname{med}_{\ell\in U}(x_{\ell3})$ denotes the finite-population median of the third auxiliary variable. Under this mechanism, the average response rate was approximately $48\%$.

\subsection*{Point estimators}

\noindent Let $S_r=\{k\in S:r_k=1\}$ denote the respondent set, and let $w_k$ denote the design weight. Under simple random sampling without replacement, $w_k=N/n$ for every sampled unit. For any positive function $q(\cdot)$, define the response-adjusted weight
\[
\omega_k(q)=\frac{w_k}{q(\mathbf{x}_k)}.
\]
Throughout this simulation, we use the Hájek-type version of the IPW estimator,
\begin{equation}\label{eq:hajek_ipw_sim}
	\widehat{\mu}_{\mathrm{ipw}}^{\mathrm{H}}(q)
	=
	\frac{\displaystyle\sum_{k\in S_r}\omega_k(q)y_k}
	{\displaystyle\sum_{k\in S_r}\omega_k(q)}.
\end{equation}
The oracle estimator is $\widehat{\mu}_{\mathrm{ipw}}^{\mathrm{H}}(p)$, which uses the true response probabilities. The feasible estimator is $\widehat{\mu}_{\mathrm{ipw}}^{\mathrm{H}}(\widehat p)$, which uses estimated response probabilities. Equation~\eqref{eq:hajek_ipw_sim} is the normalized version of the usual, unnormalized IPW estimator. The normalization can improve finite-sample stability when some estimated response probabilities are small. All point estimates and all bootstrap replicates reported below use this normalized form.

\noindent The response probabilities were estimated using the following methods:
\begin{enumerate}
	\item \textbf{Logistic regression.} A logistic regression model of $r_k$ on $(x_{k1},x_{k2},x_{k3},x_{k4})$ was fitted.
	
	\item \textbf{CART.} A regression tree was fitted to $r_k$ using the \texttt{rpart} algorithm in regression mode (\texttt{method = "anova"}), with complexity parameter \texttt{cp = 0} and minimum split size \texttt{minsplit = 20}.
	
	\item \textbf{Random forest.} A regression forest was fitted to $r_k$ using \texttt{randomForest}, with \texttt{ntree = 500}, \texttt{mtry = 2}, and \texttt{nodesize = 5}.
	
	\item \textbf{XGBoost.} Gradient boosting was applied using the \texttt{xgboost} algorithm with binary logistic loss. The main tuning parameters were \texttt{eta = 0.3}, \texttt{max\_depth = 6}, \texttt{min\_child\_weight = 1}, \texttt{subsample = 1}, \texttt{colsample\_bytree = 1}, and a maximum of $100$ boosting iterations. The final number of iterations was selected by five-fold cross-validation with early stopping.
	
	\item \textbf{Propensity score stratification (PSS).} A logistic regression model was first used to obtain estimated response scores. These scores were divided into $C=5$ strata using sample quantiles. Within stratum $c$, the response probability was estimated by the observed response rate, $\widehat p_c=n_{r,c}/n_c$, and every unit in that stratum was assigned this value. Under simple random sampling, the resulting Hájek-type estimator can also be written as
	\[
	\widehat\mu_{\mathrm{PSS}}
	=
	\sum_{c=1}^{C}\frac{n_c}{n}\,\overline y_{r,c},
	\]
	where $n_c$ is the number of sampled units in stratum $c$ and $\overline y_{r,c}$ is the respondent mean in that stratum.
\end{enumerate}

\noindent For each method, performance was assessed using Monte Carlo relative bias and root mean squared error. Let $\widehat\mu^{(1)},\ldots,\widehat\mu^{(M)}$ denote the estimates obtained over the $M=10{,}000$ Monte Carlo replications. We used
\[
\operatorname{RB}_{\mathrm{MC}}(\widehat\mu)
=
100\,\frac{\overline{\widehat\mu}_{\mathrm{MC}}-\mu}{\mu},
\qquad
\overline{\widehat\mu}_{\mathrm{MC}}
=
\frac{1}{M}\sum_{m=1}^{M}\widehat\mu^{(m)},
\]
and
\[
\operatorname{RMSE}_{\mathrm{MC}}(\widehat\mu)
=
\left\{
\frac{1}{M}\sum_{m=1}^{M}
\bigl(\widehat\mu^{(m)}-\mu\bigr)^2
\right\}^{1/2}.
\]
The RMSE values in Table~\ref{Table2} are multiplied by $100$.

\begin{table}[ht]
	\centering
	\caption{Monte Carlo relative bias (\%) and RMSE ($\times 100$) for the Hájek-type estimators.}
	\vspace{4mm}
	\label{Table2}
	\resizebox{\textwidth}{!}{
		\begin{tabular}{cc cccccc cccccc}
			\toprule
			& & \multicolumn{6}{c}{Relative bias (\%)} & \multicolumn{6}{c}{RMSE ($\times 100$)} \\
			\cmidrule(lr){3-8} \cmidrule(lr){9-14}
			$N$ & $n$
			& Oracle & Logistic & CART & RF & XGBoost & PSS
			& Oracle & Logistic & CART & RF & XGBoost & PSS \\
			\midrule
			4000  & 200
			& 0.0 & -0.1 & -0.6 & 0.2 & -1.2 & -0.2
			& 37.8 & 29.7 & 41.0 & 35.8 & 37.7 & 30.1 \\
			
			10000 & 500
			& 0.0 & 0.1 & -0.6 & 0.2 & -1.2 & -0.1
			& 23.8 & 18.7 & 26.1 & 21.8 & 26.1 & 18.7 \\
			
			20000 & 1000
			& 0.0 & 0.1 & -0.5 & 0.2 & -1.0 & -0.1
			& 16.9 & 13.1 & 18.5 & 15.2 & 19.6 & 13.2 \\
			\bottomrule
	\end{tabular}}
\end{table}

\noindent Table~\ref{Table2} shows that all methods have small relative bias in this simulation. The absolute relative bias is at most $1.2\%$ in every scenario. Logistic regression and PSS have the smallest RMSE values, while CART and XGBoost are less precise in most cases. The RMSE decreases as the sample size increases for every method. These results show that the Hájek-type estimator is nearly unbiased here, but its precision still depends on how the response probabilities are estimated.

\subsection*{Pseudo-population bootstrap}

\noindent We next evaluated a pseudo-population bootstrap estimator of the variance of $\widehat{\mu}_{\mathrm{ipw}}^{\mathrm{H}}(\widehat p)$. Pseudo-population bootstrap methods are designed to reproduce the original sampling design by drawing bootstrap samples from an artificial finite population; see, for example, Mashreghi et al.\ (2016). We focus on simple random sampling without replacement and assume that $N/n$ is an integer. Each sampled unit is then replicated exactly $N/n$ times to form a pseudo-population of size $N$.

\noindent Each copy inherits the auxiliary variables and response indicator of the corresponding sampled unit. No new response indicators are generated inside the bootstrap. Thus, whenever a copy of an original respondent is selected, it remains a respondent, and the same holds for a nonrespondent. The number and composition of respondents can still vary across bootstrap samples because different pseudo-units are selected.

\noindent For each bootstrap replication, a sample of size $n$ is drawn from the pseudo-population using simple random sampling without replacement. The response-probability method is then refitted using the bootstrap sample, and the Hájek-type estimator is recomputed. For PSS, the strata and within-stratum response rates are reconstructed in every bootstrap sample. For XGBoost, cross-validation and early stopping are repeated. 
\begin{algorithm}[htbp]
	\caption{Pseudo-population bootstrap for the Hájek-type IPW estimator}
	\label{alg:ppb_hajek}
	\begin{algorithmic}[1]
		
		\Require Original sample $S$ of size $n$, population size $N$, number of bootstrap replications $B$, and a response-probability estimation method
		\Ensure Bootstrap variance estimator $\widehat \V_{\mathrm{boot}}\!\left\{\widehat{\mu}_{\mathrm{ipw}}^{\mathrm{H}}(\widehat p)\right\}$
		
		\State Assume that $L=N/n$ is an integer.
		\State Construct a pseudo-population $U^*$ by creating $L$ copies of each unit in $S$.
		\State Let every copy inherit the original unit's auxiliary variables, response indicator, and observed outcome when $r_k=1$.
		
		\For{$b=1,\ldots,B$}
		\State Draw $S^{*(b)}$, a simple random sample without replacement of size $n$ from $U^*$.
		\State Define $S_r^{*(b)}=\{j\in S^{*(b)}:r_j^*=1\}$.
		\State Refit the same response-probability method using $\{(r_j^*,\mathbf{x}_j^*):j\in S^{*(b)}\}$ and obtain $\widehat p_j^{*(b)}$.
		\State Apply the same probability lower bound and all method-specific tuning steps used in the original sample.
		\State Set
		\[
		\omega_j^{*(b)}
		=
		\frac{w_j^{*(b)}}{\widehat p_j^{*(b)}},
		\qquad
		w_j^{*(b)}=\frac{N}{n}.
		\]
		\State Compute
		\[
		\widehat\mu_{\mathrm{ipw}}^{\mathrm{H},*(b)}
		=
		\frac{\displaystyle\sum_{j\in S_r^{*(b)}}\omega_j^{*(b)}y_j^*}
		{\displaystyle\sum_{j\in S_r^{*(b)}}\omega_j^{*(b)}}.
		\]
		\EndFor
		
		\State Compute
		\[
		\overline{\widehat\mu}_{\mathrm{ipw}}^{\mathrm{H},*}
		=
		\frac{1}{B}\sum_{b=1}^{B}
		\widehat\mu_{\mathrm{ipw}}^{\mathrm{H},*(b)}.
		\]
		
		\State Return
		\[
		\widehat \V_{\mathrm{boot}}\!\left\{\widehat\mu_{\mathrm{ipw}}^{\mathrm{H}}(\widehat p)\right\}
		=
		\frac{1}{B-1}
		\sum_{b=1}^{B}
		\left(
		\widehat\mu_{\mathrm{ipw}}^{\mathrm{H},*(b)}
		-
		\overline{\widehat\mu}_{\mathrm{ipw}}^{\mathrm{H},*}
		\right)^2.
		\]
		
	\end{algorithmic}
\end{algorithm}

\noindent The bootstrap study used $N=20{,}000$, $n=1{,}000$, $M=10{,}000$ Monte Carlo replications, and $B=250$ bootstrap samples within each Monte Carlo replication. Let $\widehat\mu_m$ and $\widehat \V_{\mathrm{boot},m}$ denote the Hájek-type point estimate and bootstrap variance estimate in Monte Carlo replication $m$. The Monte Carlo variance of the point estimator was estimated by
\[
\V_{\mathrm{MC}}(\widehat\mu)
=
\frac{1}{M-1}\sum_{m=1}^{M}
\left(\widehat\mu_m-\overline{\widehat\mu}_{\mathrm{MC}}\right)^2.
\]
The Monte Carlo percent relative bias of the bootstrap variance estimator was then defined as
\[
\operatorname{RB}_{\mathrm{MC}}(\widehat \V_{\mathrm{boot}})
=
100\,
\frac{M^{-1}\sum_{m=1}^{M}\widehat \V_{\mathrm{boot},m}
	-
	\V_{\mathrm{MC}}(\widehat\mu)}
{\V_{\mathrm{MC}}(\widehat\mu)}.
\]
We also computed the coverage probability of the Wald-type intervals
\[
\widehat\mu_m
\pm
z_{0.975}\sqrt{\widehat \V_{\mathrm{boot},m}}.
\]
The Monte Carlo coverage probability was the percentage of these intervals that contained the finite-population mean $\mu$.

\begin{table}[ht]
	\centering
	\caption{Monte Carlo percent relative bias (\%) of the bootstrap variance estimator and Monte Carlo coverage probability (\%) of the associated Wald-type confidence intervals for $N=20{,}000$ and $n=1{,}000$.}
	\label{Table3}
	\vspace{4mm}
	\begin{tabular}{lcc}
		\toprule
		Method & Relative bias (\%) & Coverage probability (\%) \\
		\midrule
		Logistic regression & -0.3 & 94.8 \\
		PSS                 & 1.6  & 94.9 \\
		CART                & 36.8 & 97.4 \\
		RF                  & 22.8 & 95.1 \\
		XGBoost             & -10.8 & 88.5 \\
		\bottomrule
	\end{tabular}
\end{table}

\noindent The results in Table~\ref{Table3} differ clearly across methods. For logistic regression and PSS, the bootstrap variance estimator has very small relative bias, and the coverage probabilities are close to the nominal $95\%$ level. For CART, the bootstrap variance is too large on average, with a relative bias of $36.8\%$, and the resulting intervals are conservative, with $97.4\%$ coverage. The random forest variance estimate is also too large on average, although its coverage remains close to $95\%$ in this setting. XGBoost shows the opposite pattern: the bootstrap variance is underestimated by about $11\%$, and the coverage probability falls to $88.5\%$. The point estimator based on XGBoost also has the largest absolute bias in Table~\ref{Table2}, which may further reduce coverage because the Wald intervals are centered at the point estimate. Overall, the pseudo-population bootstrap works well for logistic regression and PSS in this simulation, but its performance is not equally reliable for all machine learning methods.

\end{document}